\documentclass[prl, aps, twocolumn, groupedaddress,showpacs]{revtex4}
\usepackage{eqnarray}
\usepackage[english]{babel} 
\usepackage[english]{layout}
\usepackage{amsmath,amsfonts,amssymb}
\usepackage{graphicx}
\usepackage{float}
\usepackage{color}
\usepackage{braket}
\usepackage{version}
\usepackage{array,multirow,makecell}

\begin{document}

\title{Full Optical Control of Topological Transitions in Polariton Chern Insulator Analog}
\author{O.Bleu, D. D. Solnyshkov, G. Malpuech}
\affiliation{Institut Pascal, PHOTON-N2, Clermont Auvergne University, CNRS, 4 avenue Blaise Pascal, 63178 Aubi\`{e}re Cedex, France.} 

\begin{abstract}
Exciton-polariton lattices allow to implement topologically protected photonic edge states at optical frequencies. Taking advantage of the interacting character of polaritons, we show that several topological phases, belonging to the Quantum Anomalous Hall family, can occur in polariton graphene with the TE-TM spin-orbit coupling and an effective Zeeman field. This Zeeman field can be entirely controlled by the non-linearity of the macro-occupied mode created by circularly polarized resonant pumping. The topological phases found are distinct from the ones of normal graphene in presence of a combination of Rashba and Zeeman field.
\end{abstract}

\maketitle

The discovery of the quantum Hall effect \cite{PhysRevLett.45.494} and its explanation in terms of topology \cite{thouless1982quantized,PhysRevLett.51.2167} have refreshed the interest to the band theory in condensed matter physics leading to the definition of a new class of insulators \cite{RevModPhys.83.1057,RevModPhys.82.3045} including Chern Insulators with broken time reversal (TR) symmetry \cite{haldane1988model,PhysRevLett.101.146802,PhysRevB.82.161414} and Quantum Spin Hall (QSH) Topological Insulators with conserved TR  symmetry \cite{bernevig2006quantum,kane2005z,konig2007quantum}. Another field which has considerably grown these last years is the emulation of such topological insulators with fermions (either charged, as electrons in nanocrystals \cite{kalesaki2014dirac,beugeling2015topological}, or neutral, such as fermionic atoms in optical lattices \cite{jotzu2014experimental,PhysRevLett.116.225305}) and bosons (atoms, photons, or mixed light-matter quasiparticles \cite{peano2015topological}). The  main advantage of artificial analogs is the possibility to tune the parameters, to obtain inaccessible regimes, and to measure quantities out of reach in the original systems. Photonic systems have allowed the first demonstration of Quantum Hall effect without Landau levels, the Quantum Anomalous Hall effect (QAHE) \cite{Soljacic2009,Soljacic2014}, later implemented in atomic systems \cite{aidelsburger2015measuring}. They have allowed the realization of topological bands with high Chern numbers \cite{PhysRevLett.115.253901}. From an applied point of view, they open the way to non-reciprocal photonic transport, highly desirable to implement logical photonic circuits. On the other hand, the study of interacting particles in artificial topologically non-trivial bands could allow direct measurements on Laughlin wave functions \cite{PhysRevLett.108.206809} and give access to a wide variety of strongly interacting fermionic \cite{wang2014classification} and bosonic phases \cite{peotta2015superfluidity}. In that framework, the recent proposals of implementing topological insulator analogs for a mixed exciton-photon particle (polariton) \cite{nalitov2015polariton,KarzigPRX2015,LiewPRB2015,PhysRevB.93.104303,PhysRevB.93.020502,PhysRevB.93.085438}, combining the flexibility of design and study of photonic systems with the behavior of an interacting spinor quantum fluid \cite{carusotto2013quantum,shelykh2009polariton}, have a strong fundamental and applied interest.

 One of the key requirements to implement topological insulator analogs for neutral particles is the creation of either synthetic magnetic fields or a spin-orbit coupling (SOC), a milestone achieved a few years ago in atomic systems \cite{lin2009synthetic,lin2011spin}. In photonics, several schemes have been proposed and realized, many of them based on a specific lattice design inducing complex hopping from one cite to the other \cite{rechtsman2013photonic,hafezi2011robust,hafezi2013imaging,khanikaev2013photonic}. However, the initial Haldane-Raghu proposal \cite{haldane2008possible}, demonstrated experimentally in several works \cite{Soljacic2009,Soljacic2014,PhysRevLett.115.253901} does not rely on such structural design, but rather on a standard honeycomb or square lattice with  band degeneracies at the corners of the Brillouin zone, in which the TR symmetry is broken by the Faraday effect (equivalent to Zeeman splitting). This requires the use of materials with a good magnetic response and limits the experimental realizations to the microwave range. We point out that a Zeeman field alone is not sufficient and that the second key ingredient is the use of the TE and TM nature of the photonic modes, which makes a non-Abelian gauge field for photons \cite{PhysRevLett.114.026803,PhysRevB.93.085438} emerge at the corners of the Brillouin zone. The two fields combined give a vortical effective field texture which opens a gap with same-sign Berry curvature at all corners of the Brillouin zone \cite{PhysRevLett.102.046407}.

This combination has been used in the proposals for  polariton Chern insulators \cite{nalitov2015polariton,KarzigPRX2015,LiewPRB2015}. The polaritons inherit their Zeeman splitting under magnetic field from the excitonic fraction, which provides a good magnetic activity at optical frequencies \cite{PhysRevLett.112.093902,PhysRevB.91.155130}. The splitting between the TE and TM polariton modes in planar cavities is  well known as a SOC responsible for the optical spin Hall effect \cite{PhysRevLett.95.136601, leyder2007observation} and the generation of topologically protected spin currents in polaritonic molecules \cite{PhysRevX.5.011034}. The relatively weak TE-TM splitting in planar cavities leads to Chern numbers equal to $\mp 2$, contrary to photonic crystals, where the Chern numbers are typically $\pm 1$. Polaritons also behave as a quantum fluid of light \cite{carusotto2013quantum}.  A dense polariton gas with a well defined velocity, phase, and polarization can be resonantly driven by an external laser pump which allows the creation of superfluid flows \cite{PhysRevLett.93.166401} and generation of half-integer topological defects \cite{hivet2012half,dominici2014vortex} in these flows. The interactions in such gas are strongly spin-anisotropic \cite{shelykh2009polariton} which creates a self-induced Zeeman (SIZ) field in case of a spin population imbalance.

In this manuscript, we demonstrate the full optical control of the band topology in a resonantly driven photonic (polaritonic) lattice, without any applied magnetic field.  We show that the topologically trivial band structure becomes non-trivial under resonant circularly polarized pumping at the $\Gamma$ point of the dispersion. A self-induced topological gap opens in the dispersion of the elementary excitations. We show that the tuning of the pump intensity allows to go through several topological transitions. We establish a complete phase diagram for this system, demonstrating the inversion of chirality.  We compare the case of TE-TM SOC (polariton graphene \cite{PhysRevLett.112.116402}) with the Rashba SOC, relevant for atomic lattices and previously considered for normal graphene \cite{PhysRevB.82.161414,zhang2015quantum}, analyzing the differences between the phase diagrams.
\begin{figure}[tbp]
 \begin{center}
 \includegraphics[scale=0.46]{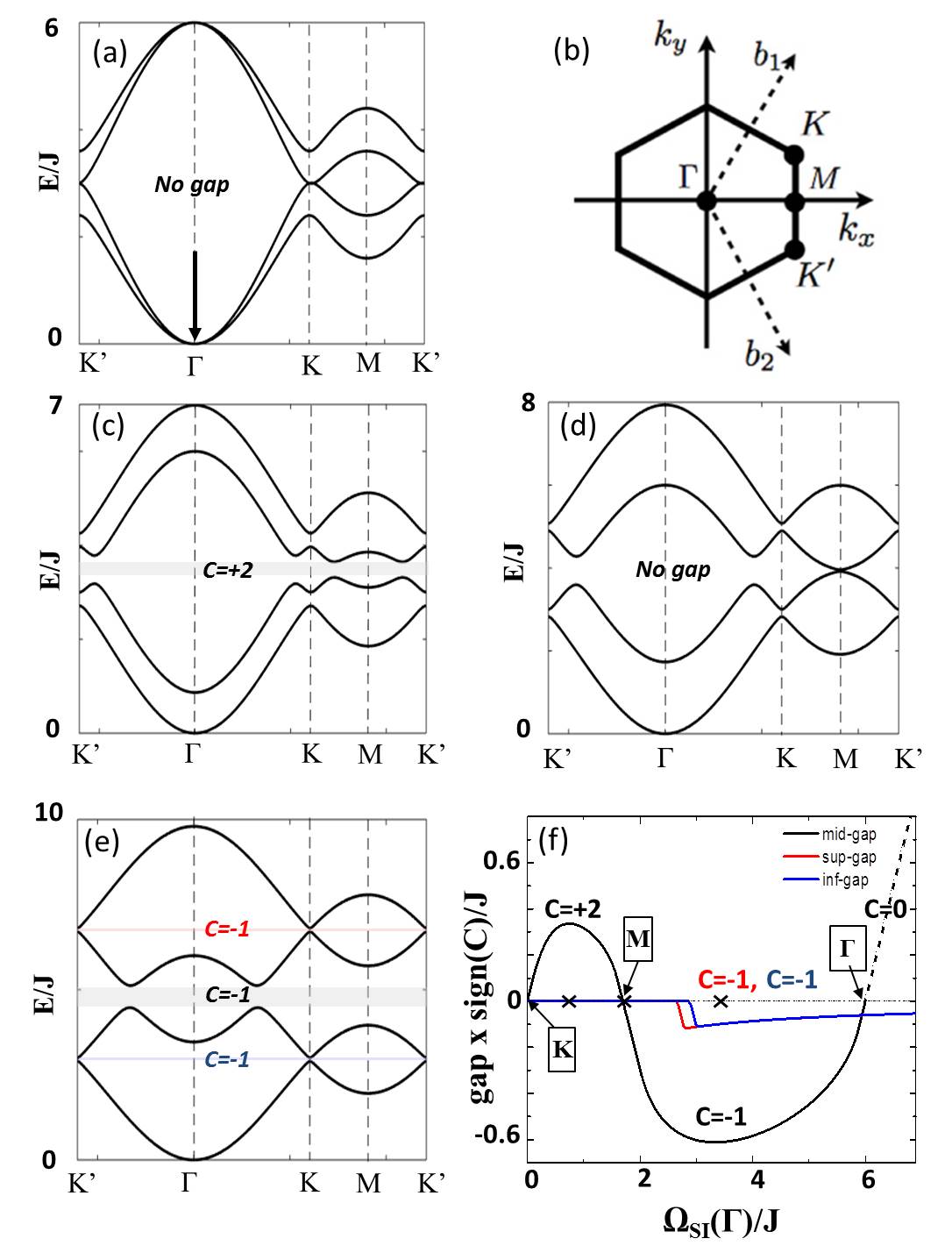}
 \caption{(a) Polariton band structure in the honeycomb lattice. The arrow represents the laser pump. (b) First Brillouin zone with the high symmetry points. (c,d,e) elementary excitation dispersions for different topological phases $\alpha_1n/J=1$ , $\alpha_1n/J=2$, and $\alpha_1n/J=4$ with the corresponding gap Chern numbers. (f) The gaps as functions of $\Omega_{SI}$ and their Chern numbers. The three crosses correspond to the band structures (c), (d), and (e). The high symmetry points where the middle gap closes are written ($J=1,\delta J=0.2 J$). }
  \end{center}
 \end{figure}

 We begin our consideration with the tight-binding Hamiltonian of polariton graphene: 
\begin{eqnarray}
H_k=\begin{pmatrix}
0 &F_k \\
F_k^+& 0
\end{pmatrix} 
,\quad F_k=-\begin{pmatrix}
f_kJ&f_k^+\delta J \\
f_k^- \delta J&f_kJ 
\end{pmatrix}
\end{eqnarray}
with the complex coefficients $f_k=\sum_{j=1}^3\exp{(-i\textbf{kd}_{\phi_j})}$ and $f_k^{\pm}=\sum_{j=1}^3 \exp{(-i[\textbf{kd}_{\phi_j}\mp 2\phi_j])}$.
This Hamiltonian is a 4-by-4 matrix written in the basis of the sublattice (A/B) pseudospin and of the polarization (+/-) pseudospin. $J$ is the tunneling coefficient between nearest neighbour pillars (A/B) and $\delta J$ is the TE-TM SOC (+/-). The TE-TM splitting for guided modes in photonic stuctures is typically quite large, only a specific combination of parameters allows to cancel it locally \cite{khanikaev2013photonic,cheng2016robust}. It is much weaker for the radiative modes of a planar cavities that we consider here, but not negligible, having led to the observation of a very wide variety of effects in the last decade \cite{shelykh2009polariton,leyder2007observation}. A typical dispersion is shown in Fig.~1(a), and the first Brillouin zone with high symmetry points is shown in Fig.~1(b).

A coherent macro-occupied state of exciton-polaritons is usually created by resonant optical excitation. This regime is well described in the mean-field approximation \cite{PhysRevLett.93.166401,PhysRevB.77.045314}. We can derive the driven tight-binding  Gross-Pitaevskii equation in this honeycomb lattice for a homogeneous laser pump $F$ ($\hbar=1$).
\begin{equation}
i\frac{\partial}{\partial t}\Psi_s^\sigma=[H_s^\sigma(\textbf{k})+\alpha_1|\Psi_s^\sigma|^2+\alpha_2|\Psi_s^{-\sigma}|^2]\Psi_s^\sigma+ F_{s}^{\sigma}e^{i(\textbf{(k                          }_p.\textbf{r}-\omega_pt)}
\label{GP-graphene}
\end{equation}
$s$ is the sublattice index and $\sigma$ is the spin index. $H_s^\sigma$ is the tight-binding Hamiltonian defined above. $\alpha_1$ and $\alpha_2$ are the interaction constants between particles with the same and opposite spins respectively. For polaritons, the latter is usually much weaker: $ |\alpha_2| \ll \alpha_1 $ \cite{Renucci2005,Vladimirova2010}, and we neglect it.  $F_{s}^{\sigma}$ is the pump amplitude. In the following, we consider a homogeneous pump, which implies that its amplitude on A and B pillars is the same. However, the spin projections $F_A^+$ and $F_A^-$, determining the polarization of the pump, can be different. The quasi-stationary driven solution has the same frequency and wavevector as the pump ($\Psi_{s}^{\sigma}=e^{i(\textbf{k}_p.\textbf{r}-\omega_pt)}\Psi_{p,s}^{\sigma}$). Its wavefunction satisfies a set equations:
\begin{eqnarray}
&(\omega_p + i \gamma_p -\alpha_1 |\Psi_{p,s}^{\sigma}|^2 -\alpha_2 |\Psi_{p,s}^{-\sigma}|^2) \Psi_{p,s}^{\sigma} \nonumber \\
&+f_{k_p}J \Psi_{p,-s}^{\sigma} +f_{k_p}^{\sigma}J \Psi_{p,-s}^{-\sigma}=F_s^{\sigma} \label{stat1}
\end{eqnarray}
where $\omega_p$ is the frequency of the pump mode and $\gamma_p$ is the linewidth. The tight-binding terms $f_{k_p}$ of the polariton graphene induce a coupling between the sublattices and polarizations. Note that Eq. \eqref{stat1} is written for an arbitrary pump wave vector $k_p$. In the following, we consider a pump, resonant with the energy of the lower polariton dispersion branch in the $\Gamma$ point ($\omega_p=-f_{\Gamma}J=-3J$ and $k_p=0$), marked with an arrow in Fig.~1(a).
We compute the dispersion of the elementary excitations using the standard wave function of a weak perturbation ($|\textbf{u}|$,$|\textbf{v}|\ll|\vec{\Phi}_p |$):
\begin{equation}
\vec{\Phi}=e^{i(\textbf{k}_p.r-\omega_p t)}(\vec{\Phi}_p+\textbf{u}e^{i(\textbf{k}.r-\omega t)}+\textbf{v}^*e^{-i(\textbf{k}.r-\omega^* t)})
\end{equation}
where $\vec{\Phi}_p=(\Psi_{p,A}^+,\Psi_{p,A}^-,\Psi_{p,B}^+,\Psi_{p,B}^-)^T$ , $\textbf{u}$ and $\textbf{v}$ are vectors of the form $(u_A^+,u_A^-,u_B^+,u_B^-)^T$ (details are given in the supplementary material \cite{suppl}).
 
If the pump is totally circularly polarized, the macro-occupied state is also circularly polarized ($n^-=0$), and $n=n^+=n_A^+ +n_B^+=|\Psi_{p,A}^{+}|^2+|\Psi_{p,B}^{+}|^2
$. A SIZ splitting appears due to the spin-anisotropic interactions and to the difference of populations of the two spin components. A simple analytical formula of the $k$-dependent SIZ splitting between the two lower branches of the graphene dispersion is obtained by neglecting the SOC ($\delta J =0$): 
\begin{equation}
\Omega_{SI}=\omega_p+|f_k|+\sqrt{(\omega_p+|f_k|J-2 \alpha_1 n_{A/B}^+)^2- (\alpha_1 n_{A/B}^+)^2}\nonumber
\end{equation}
One of the key difference with respect to the Zeeman field created by an applied magnetic field is the  dependence of the SIZ field on the wave vector and energies of the bare modes. This dependence has already been shown to lead to the inversion of the effective field sign at the K point (and thus the inversion of the topology) when both applied and SIZ fields created by a Bose-Einstein condensate are present \cite{PhysRevB.93.085438}.

Figure 1(c)-(e) shows the weak excitation dispersion for three different values of the SIZ field $\Omega_{SI}$ at \textbf{k}=0 (taken as a reference for comparison, because the $\Gamma$ point is not affected by the SOC), whereas the Fig. 1(f) shows the magnitude of the different gaps multiplied by the sign of the Chern number of the valence band ($C=\sum_{i=1}^n C_n$) \cite{PhysRevLett.71.3697} versus the same SIZ field at \textbf{k}=0. This SIZ field, which is the blueshift of the "+" spin component, can be calculated as $\Omega_{SI}(\Gamma)=\sqrt{3}\alpha_1n/2$   \cite{suppl}. When its value is between 0 and the first transition point (approximately $2J$, Fig. 1(f)), a single gap opens at the K points and the valence band Chern number is +2 (Fig. 1(c)). The gap value at K increases versus $\alpha_1n$, but it decreases at the M point, where the gap closes when $\sqrt{\Omega_{SI}^2(k_M)+4\delta J^2}=J$, as shown in Fig. 1(d). Increasing the pumping further leads to an immediate reopening of the gap at the M point with the Chern numbers changing values and sign from $\mp 2$ to $\pm 1$. When the pumping is increased further and $\Omega_{SI}$ becomes of the order of $3J$, two additional topological gaps open at the K points, in the middle of the conduction and valence bands respectively (Fig. 1(e)), because the $\Gamma$ point of the upper branch rises above the K point of the lower branch (of the same band). These two new gaps do not open at the same pumping value because of the difference between the SIZ fields in the upper and lower bands.
One should notice that in this pumping range the main gap reaches its largest value, of the order of $0.6J$, which is about $3 \delta J$. In realistic samples, this means a gap of 0.2-1 meV, quite larger than the linewidth \cite{PhysRevLett.112.116402}. Increasing the pumping further closes the gap at the $\Gamma$ point, and then the middle gap becomes trivial.

In order to visualize the dispersion of the surface states, we compute the eigenmodes of a graphene ribbon for two different values of $\alpha_1n$ (Fig. 2(a),(b)) corresponding to the values used in Figs. 1(c) and 1(e) respectively. In Fig. 2(a), there is only one topological gap characterized by a Chern number $+2$ and hence there are two edge modes on each side of the ribbon. In Fig. 2(b), we can observe three topological gaps with the Chern number of the top and bottom bands being $\pm 1$ respectively. Each of them is characterized by the presence of only one edge mode on a given edge of the ribbon, and the group velocities of the modes are opposite to the previous phase: the chirality is controlled by the intensity of the pump. This inversion, associated with the change of the topological phase ($\left| C \right| = 2 \to 1$), is fundamentally different from the one of Ref. \cite{PhysRevB.93.085438}, observed for the same phase ($\left|C\right|=2$).
 \begin{figure}[t]
 \begin{center}
 \includegraphics[scale=0.7]{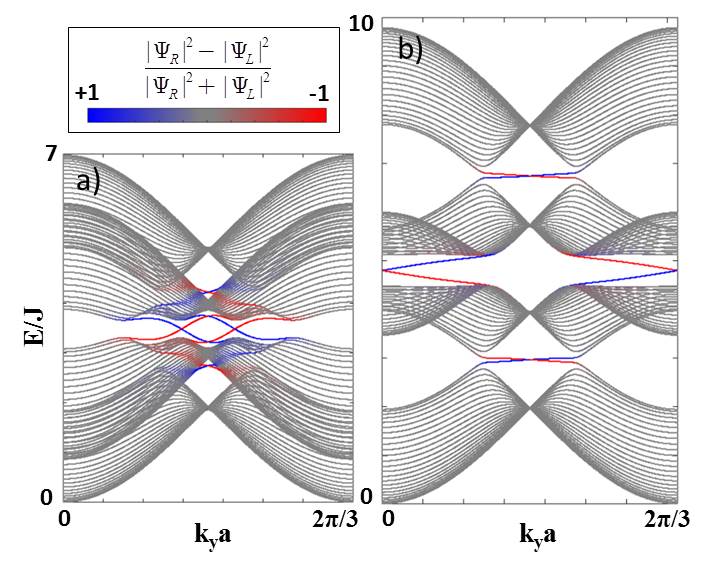}
 \caption{ (Color online) Band structure of a graphene ribbon in two different phases. Blue and red colors refer to the states localized on the right and left edges. Parameters:  $\delta J=0.2J$ and (a) $\alpha_1n=1J$, (b) $\alpha_1n=4J$.}
  \end{center}
 \end{figure}
Figure 3(a) shows the complete topological phase diagram versus the SIZ splitting $\Omega_{SI}$ in the $\Gamma$ point and $\delta J$. Each phase is marked by the Chern numbers of the corresponding bands. One should notice that to change the SOC ($\delta J$) one needs to change the structure, while the SIZ field $\Omega_{SI}$ can be directly modulated by the pumping intensity. We observe that the closing of the gap at the M point can occur by increasing either $\delta J$ or $\Omega_{SI}$. The limit $\delta J/J\approx 1$ is typical for photonic crystal slabs with large TE-TM SOC, and  the topological phase we find with Chern numbers equal to $\pm 1$ is qualitatively similar with the Haldane-Raghu phase \cite{haldane2008possible}. A phase diagram similar to Fig.~3(a) can be also obtained by varying an external magnetic field (see \cite{suppl}). A qualitative difference between the two situations (SIZ and applied magnetic field) is that the two extra gaps which appear in the valence and conduction bands do not open at the same value of $\Omega$, because of the energy dependence of $\Omega_{SI}$ in the resonant pumping case.
 \begin{figure}[t]
 \begin{center}
 \includegraphics[scale=0.37]{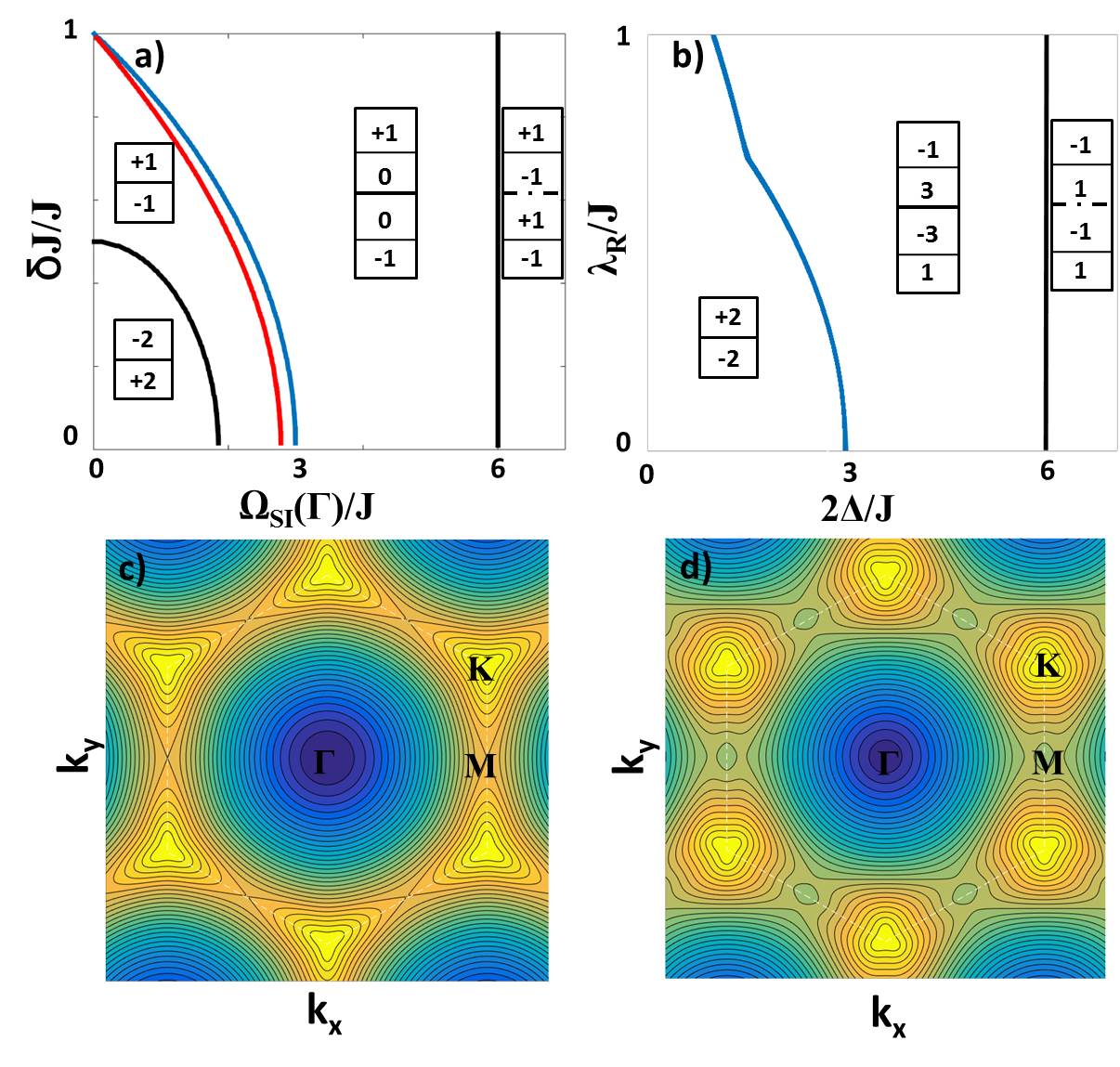}
 \caption{ (Color online) Phase diagrams (a) for the TE-TM SOC and $\Omega_{SI}$ (resonant pumping); (b) for the Rashba SOC and applied field $\Delta$. Each phase is marked by the Chern numbers of the bands. The red and blue curves correspond to the opening of additional gaps. (c)-(d) Graphene bare dispersions with TE-TM and Rashba SOC respectively (2nd branch) ($\delta J=\lambda_R=0.2J$).}
  \label{disp}
  \end{center}
 \end{figure}

We have also calculated a phase diagram for a honeycomb lattice taking into account Rashba SOC (see \cite{suppl}), shown in Fig.~3(b).  The two phase diagrams are qualitatively different. The first transition of the TE-TM SOC is absent for the Rashba SOC, because the trigonal warping of the bare dispersion around K points occurs in different directions, as shown in Fig.~3(c),(d). In the TE-TM case, the additional Dirac points appear in the $K-M$ directions, allowing them to merge at the $M$ point and disappear (Fig.~1(d)). In the Rashba configuration, the additional Dirac points in the $K-\Gamma$ direction cannot merge, and the only remaining transitions are $2\Delta=3J$ (branch splitting) and $2\Delta=6J$ (central gap becomes trivial, its edge states disappear). 

 One should note that even if the two single particle Hamiltonians allow some comparison, for instance trigonal warping is present in both cases, their physical meanings are different. Honeycomb lattice with Rashba SOC is a typical situation for electrons in graphene \cite{PhysRevB.79.165442}, where the Zeeman term, leading to the  QAHE, appears generally due to exchange-type interactions \cite{PhysRevB.82.161414} and the main tunable parameter is the Rashba SOC $\lambda_R$. Moreover, to our knowledge, despite several theoretical proposals \cite{PhysRevB.82.161414,zhang2015quantum}, the QAHE has not been observed experimentally in graphene yet. 
In our case, the single particle Hamiltonian describes the artificial polaritonic graphene structure made of coupled optical cavities, and we demonstrate that this model can be extended to describe the weak excitations of a macro-occupied state and hence allow a full optical control of the topologically protected edge states of light.  In recent theoretical works \cite{PhysRevB.93.085438,peano2015topological}, topologically protected Bogoliubov edge modes have also been found in two different driven schemes based on Kagome lattices. Both models neglect the polarization degree of freedom, and, as a consequence, obtain trivial hopping in the linear case. Their topological gap is opened by the bogolon creation mechanism. In our model, the linear band structure is already characterized by non trivial hopping, and the gap is opened by the combination of the SOC and the effective Zeeman field due to spin population imbalance.
 \begin{figure}[tbp]
 \begin{center}
 \includegraphics[scale=0.32]{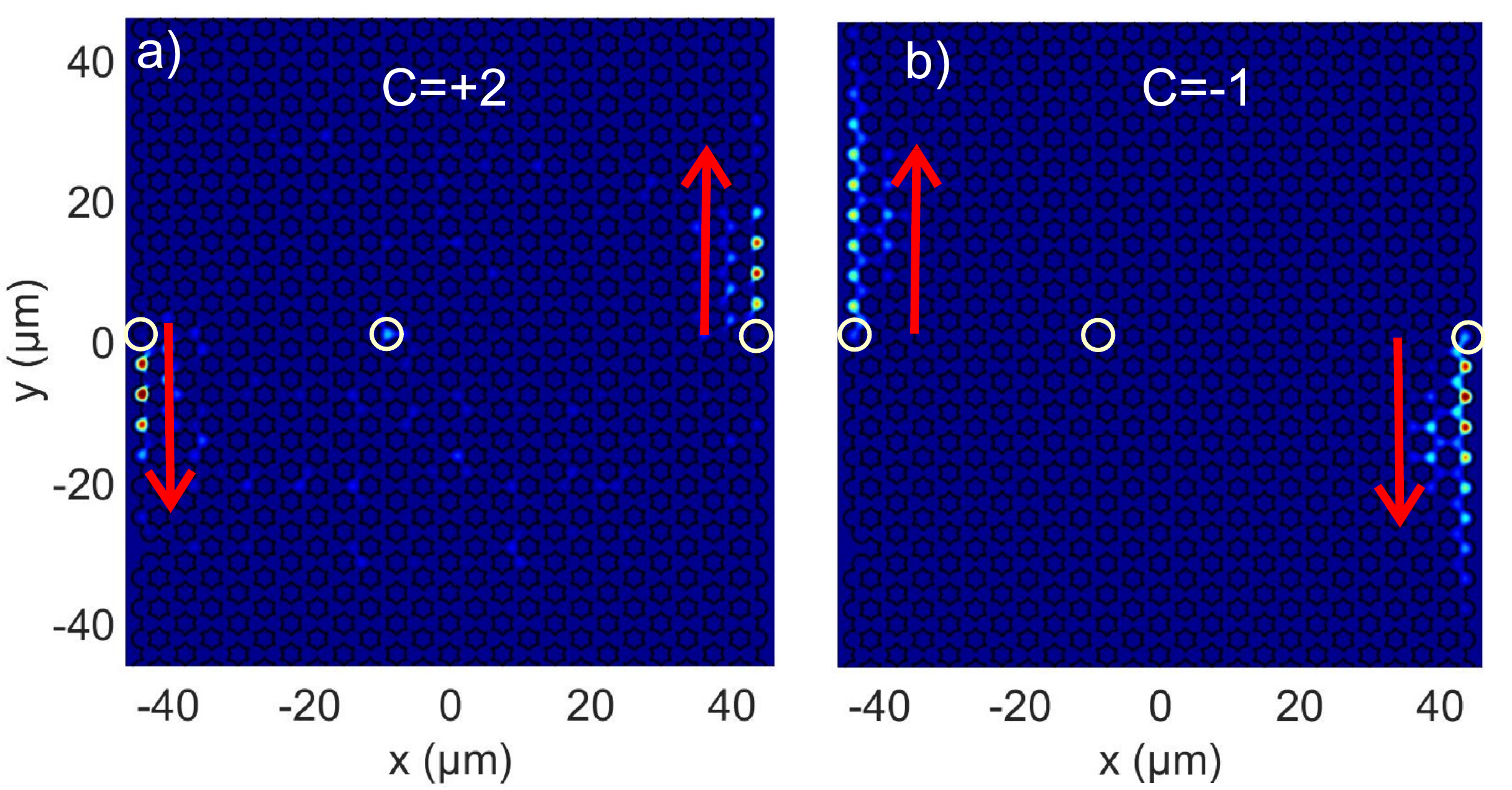}
 \caption{ (Color online) Calculated images of emission $\left|\psi_+\right|^2+\left|\psi_-\right|^2$ from the surface states (a) $\Omega_{SI}=0.3$ meV, $C=2$  (b) $\Omega_{SI}=0.6$ meV, C=-1. Arrows mark the propagation direction. }
  \label{surfacestates}
  \end{center}
 \end{figure}

To confirm our analytical predictions and to check the possibility of their experimental observation, we perform a full numerical simulation without the tight-binding or Bogoliubov approximations. We solve the spinor Gross-Pitaevskii equation for polaritons with quasi-resonant pumping with a honeycomb lattice potential $U$:

\begin{eqnarray}
& i\hbar \frac{{\partial \psi _ \pm  }}
{{\partial t}}  =  - \frac{{\hbar ^2 }}
{{2m}}\Delta \psi _ \pm +\alpha_1\left|\psi_\pm\right|^2\psi_\pm  + U\psi _ \pm   - \frac{{i\hbar }}
{{2\tau }}\psi _ \pm  \\
& + \beta {\left( {\frac{\partial }{{\partial x}} \mp i\frac{\partial }{{\partial y}}} \right)^2}{\psi _ \mp } + P_{0+}e^{-i\omega t} \notag \\ 
& +\sum_{j}P_{j-} e^{ { - \frac{{\left( {t - t_0 } \right)^2 }}
{{\tau _0^2 }}}}e^{ { - \frac{{\left( {{\mathbf{r}} - {\mathbf{r}}_j } \right)^2 }}
{{\sigma ^2 }}}}e^{ -{i\omega t} } \notag
\end{eqnarray}
where ${\psi_+(\mathbf{r},t), \psi_-(\mathbf{r},t)}$ are the two circular components of the wave function, $m=5\times10^{-5}m_{el}$ is the polariton mass, $\tau=30$ ps the lifetime. The main pumping term $P_{0+}$ is circular polarized ($\sigma^+$) and spatially homogeneous, while the 3 pulsed probes are $\sigma^-$ and localized on 3 pillars: 2 on the edges and 1 in the bulk. Results (filtered at the energy of the edge states) are shown in Fig. \ref{surfacestates}. As compared with the previously analyzed\cite{nalitov2015polariton,PhysRevB.93.085438} $C=2$ case (a), a larger gap of the $C=-1$ phase (b) demonstrates a better protection against bulk excitation, a longer propagation distance, and an inverted propagation direction, all achieved by changing the intensity of optical pumping.

\emph{Conclusions}. We have analyzed the elementary excitations of a honeycomb lattice of micro-pillars with TE-TM SOC, pumped homogeneously at \textbf{k}=0. We have demonstrated that the circularity of the laser implies the opening of a topological gap without any external magnetic field. Different phases characterized by topological invariants $\mp 2$ and $\pm1$ are achievable. The topological phase transitions can be observed by varying the pump power, which is the main tunable parameter in experiments.

 \bibliography{biblio} 

\begin{thebibliography}{58}
\expandafter\ifx\csname natexlab\endcsname\relax\def\natexlab#1{#1}\fi
\expandafter\ifx\csname bibnamefont\endcsname\relax
  \def\bibnamefont#1{#1}\fi
\expandafter\ifx\csname bibfnamefont\endcsname\relax
  \def\bibfnamefont#1{#1}\fi
\expandafter\ifx\csname citenamefont\endcsname\relax
  \def\citenamefont#1{#1}\fi
\expandafter\ifx\csname url\endcsname\relax
  \def\url#1{\texttt{#1}}\fi
\expandafter\ifx\csname urlprefix\endcsname\relax\def\urlprefix{URL }\fi
\providecommand{\bibinfo}[2]{#2}
\providecommand{\eprint}[2][]{\url{#2}}

\bibitem[{\citenamefont{Klitzing et~al.}(1980)\citenamefont{Klitzing, Dorda,
  and Pepper}}]{PhysRevLett.45.494}
\bibinfo{author}{\bibfnamefont{K.~v.} \bibnamefont{Klitzing}},
  \bibinfo{author}{\bibfnamefont{G.}~\bibnamefont{Dorda}}, \bibnamefont{and}
  \bibinfo{author}{\bibfnamefont{M.}~\bibnamefont{Pepper}},
  \bibinfo{journal}{Phys. Rev. Lett.} \textbf{\bibinfo{volume}{45}},
  \bibinfo{pages}{494} (\bibinfo{year}{1980}),
  \urlprefix\url{http://link.aps.org/doi/10.1103/PhysRevLett.45.494}.

\bibitem[{\citenamefont{Thouless et~al.}(1982)\citenamefont{Thouless, Kohmoto,
  Nightingale, and Den~Nijs}}]{thouless1982quantized}
\bibinfo{author}{\bibfnamefont{D.}~\bibnamefont{Thouless}},
  \bibinfo{author}{\bibfnamefont{M.}~\bibnamefont{Kohmoto}},
  \bibinfo{author}{\bibfnamefont{M.}~\bibnamefont{Nightingale}},
  \bibnamefont{and} \bibinfo{author}{\bibfnamefont{M.}~\bibnamefont{Den~Nijs}},
  \bibinfo{journal}{Physical Review Letters} \textbf{\bibinfo{volume}{49}},
  \bibinfo{pages}{405} (\bibinfo{year}{1982}).

\bibitem[{\citenamefont{Simon}(1983)}]{PhysRevLett.51.2167}
\bibinfo{author}{\bibfnamefont{B.}~\bibnamefont{Simon}},
  \bibinfo{journal}{Phys. Rev. Lett.} \textbf{\bibinfo{volume}{51}},
  \bibinfo{pages}{2167} (\bibinfo{year}{1983}),
  \urlprefix\url{http://link.aps.org/doi/10.1103/PhysRevLett.51.2167}.

\bibitem[{\citenamefont{Qi and Zhang}(2011)}]{RevModPhys.83.1057}
\bibinfo{author}{\bibfnamefont{X.-L.} \bibnamefont{Qi}} \bibnamefont{and}
  \bibinfo{author}{\bibfnamefont{S.-C.} \bibnamefont{Zhang}},
  \bibinfo{journal}{Rev. Mod. Phys.} \textbf{\bibinfo{volume}{83}},
  \bibinfo{pages}{1057} (\bibinfo{year}{2011}),
  \urlprefix\url{http://link.aps.org/doi/10.1103/RevModPhys.83.1057}.

\bibitem[{\citenamefont{Hasan and Kane}(2010)}]{RevModPhys.82.3045}
\bibinfo{author}{\bibfnamefont{M.~Z.} \bibnamefont{Hasan}} \bibnamefont{and}
  \bibinfo{author}{\bibfnamefont{C.~L.} \bibnamefont{Kane}},
  \bibinfo{journal}{Rev. Mod. Phys.} \textbf{\bibinfo{volume}{82}},
  \bibinfo{pages}{3045} (\bibinfo{year}{2010}),
  \urlprefix\url{http://link.aps.org/doi/10.1103/RevModPhys.82.3045}.

\bibitem[{\citenamefont{Haldane}(1988)}]{haldane1988model}
\bibinfo{author}{\bibfnamefont{F.~D.~M.} \bibnamefont{Haldane}},
  \bibinfo{journal}{Physical Review Letters} \textbf{\bibinfo{volume}{61}},
  \bibinfo{pages}{2015} (\bibinfo{year}{1988}).

\bibitem[{\citenamefont{Liu et~al.}(2008)\citenamefont{Liu, Qi, Dai, Fang, and
  Zhang}}]{PhysRevLett.101.146802}
\bibinfo{author}{\bibfnamefont{C.-X.} \bibnamefont{Liu}},
  \bibinfo{author}{\bibfnamefont{X.-L.} \bibnamefont{Qi}},
  \bibinfo{author}{\bibfnamefont{X.}~\bibnamefont{Dai}},
  \bibinfo{author}{\bibfnamefont{Z.}~\bibnamefont{Fang}}, \bibnamefont{and}
  \bibinfo{author}{\bibfnamefont{S.-C.} \bibnamefont{Zhang}},
  \bibinfo{journal}{Phys. Rev. Lett.} \textbf{\bibinfo{volume}{101}},
  \bibinfo{pages}{146802} (\bibinfo{year}{2008}),
  \urlprefix\url{http://link.aps.org/doi/10.1103/PhysRevLett.101.146802}.

\bibitem[{\citenamefont{Qiao et~al.}(2010)\citenamefont{Qiao, Yang, Feng, Tse,
  Ding, Yao, Wang, and Niu}}]{PhysRevB.82.161414}
\bibinfo{author}{\bibfnamefont{Z.}~\bibnamefont{Qiao}},
  \bibinfo{author}{\bibfnamefont{S.~A.} \bibnamefont{Yang}},
  \bibinfo{author}{\bibfnamefont{W.}~\bibnamefont{Feng}},
  \bibinfo{author}{\bibfnamefont{W.-K.} \bibnamefont{Tse}},
  \bibinfo{author}{\bibfnamefont{J.}~\bibnamefont{Ding}},
  \bibinfo{author}{\bibfnamefont{Y.}~\bibnamefont{Yao}},
  \bibinfo{author}{\bibfnamefont{J.}~\bibnamefont{Wang}}, \bibnamefont{and}
  \bibinfo{author}{\bibfnamefont{Q.}~\bibnamefont{Niu}},
  \bibinfo{journal}{Phys. Rev. B} \textbf{\bibinfo{volume}{82}},
  \bibinfo{pages}{161414} (\bibinfo{year}{2010}),
  \urlprefix\url{http://link.aps.org/doi/10.1103/PhysRevB.82.161414}.

\bibitem[{\citenamefont{Bernevig et~al.}(2006)\citenamefont{Bernevig, Hughes,
  and Zhang}}]{bernevig2006quantum}
\bibinfo{author}{\bibfnamefont{B.~A.} \bibnamefont{Bernevig}},
  \bibinfo{author}{\bibfnamefont{T.~L.} \bibnamefont{Hughes}},
  \bibnamefont{and} \bibinfo{author}{\bibfnamefont{S.-C.} \bibnamefont{Zhang}},
  \bibinfo{journal}{Science} \textbf{\bibinfo{volume}{314}},
  \bibinfo{pages}{1757} (\bibinfo{year}{2006}).

\bibitem[{\citenamefont{Kane and Mele}(2005)}]{kane2005z}
\bibinfo{author}{\bibfnamefont{C.~L.} \bibnamefont{Kane}} \bibnamefont{and}
  \bibinfo{author}{\bibfnamefont{E.~J.} \bibnamefont{Mele}},
  \bibinfo{journal}{Physical review letters} \textbf{\bibinfo{volume}{95}},
  \bibinfo{pages}{146802} (\bibinfo{year}{2005}).

\bibitem[{\citenamefont{K{\"o}nig et~al.}(2007)\citenamefont{K{\"o}nig,
  Wiedmann, Br{\"u}ne, Roth, Buhmann, Molenkamp, Qi, and
  Zhang}}]{konig2007quantum}
\bibinfo{author}{\bibfnamefont{M.}~\bibnamefont{K{\"o}nig}},
  \bibinfo{author}{\bibfnamefont{S.}~\bibnamefont{Wiedmann}},
  \bibinfo{author}{\bibfnamefont{C.}~\bibnamefont{Br{\"u}ne}},
  \bibinfo{author}{\bibfnamefont{A.}~\bibnamefont{Roth}},
  \bibinfo{author}{\bibfnamefont{H.}~\bibnamefont{Buhmann}},
  \bibinfo{author}{\bibfnamefont{L.~W.} \bibnamefont{Molenkamp}},
  \bibinfo{author}{\bibfnamefont{X.-L.} \bibnamefont{Qi}}, \bibnamefont{and}
  \bibinfo{author}{\bibfnamefont{S.-C.} \bibnamefont{Zhang}},
  \bibinfo{journal}{Science} \textbf{\bibinfo{volume}{318}},
  \bibinfo{pages}{766} (\bibinfo{year}{2007}).

\bibitem[{\citenamefont{Kalesaki et~al.}(2014)\citenamefont{Kalesaki, Delerue,
  Smith, Beugeling, Allan, and Vanmaekelbergh}}]{kalesaki2014dirac}
\bibinfo{author}{\bibfnamefont{E.}~\bibnamefont{Kalesaki}},
  \bibinfo{author}{\bibfnamefont{C.}~\bibnamefont{Delerue}},
  \bibinfo{author}{\bibfnamefont{C.~M.} \bibnamefont{Smith}},
  \bibinfo{author}{\bibfnamefont{W.}~\bibnamefont{Beugeling}},
  \bibinfo{author}{\bibfnamefont{G.}~\bibnamefont{Allan}}, \bibnamefont{and}
  \bibinfo{author}{\bibfnamefont{D.}~\bibnamefont{Vanmaekelbergh}},
  \bibinfo{journal}{Physical Review X} \textbf{\bibinfo{volume}{4}},
  \bibinfo{pages}{011010} (\bibinfo{year}{2014}).

\bibitem[{\citenamefont{Beugeling et~al.}(2015)\citenamefont{Beugeling,
  Kalesaki, Delerue, Niquet, Vanmaekelbergh, and
  Smith}}]{beugeling2015topological}
\bibinfo{author}{\bibfnamefont{W.}~\bibnamefont{Beugeling}},
  \bibinfo{author}{\bibfnamefont{E.}~\bibnamefont{Kalesaki}},
  \bibinfo{author}{\bibfnamefont{C.}~\bibnamefont{Delerue}},
  \bibinfo{author}{\bibfnamefont{Y.-M.} \bibnamefont{Niquet}},
  \bibinfo{author}{\bibfnamefont{D.}~\bibnamefont{Vanmaekelbergh}},
  \bibnamefont{and} \bibinfo{author}{\bibfnamefont{C.~M.} \bibnamefont{Smith}},
  \bibinfo{journal}{Nature communications} \textbf{\bibinfo{volume}{6}},
  \bibinfo{pages}{6316} (\bibinfo{year}{2015}).

\bibitem[{\citenamefont{Jotzu et~al.}(2014)\citenamefont{Jotzu, Messer,
  Desbuquois, Lebrat, Uehlinger, Greif, and Esslinger}}]{jotzu2014experimental}
\bibinfo{author}{\bibfnamefont{G.}~\bibnamefont{Jotzu}},
  \bibinfo{author}{\bibfnamefont{M.}~\bibnamefont{Messer}},
  \bibinfo{author}{\bibfnamefont{R.}~\bibnamefont{Desbuquois}},
  \bibinfo{author}{\bibfnamefont{M.}~\bibnamefont{Lebrat}},
  \bibinfo{author}{\bibfnamefont{T.}~\bibnamefont{Uehlinger}},
  \bibinfo{author}{\bibfnamefont{D.}~\bibnamefont{Greif}}, \bibnamefont{and}
  \bibinfo{author}{\bibfnamefont{T.}~\bibnamefont{Esslinger}},
  \bibinfo{journal}{Nature} \textbf{\bibinfo{volume}{515}},
  \bibinfo{pages}{237} (\bibinfo{year}{2014}).

\bibitem[{\citenamefont{Vanhala et~al.}(2016)\citenamefont{Vanhala, Siro,
  Liang, Troyer, Harju, and T\"orm\"a}}]{PhysRevLett.116.225305}
\bibinfo{author}{\bibfnamefont{T.~I.} \bibnamefont{Vanhala}},
  \bibinfo{author}{\bibfnamefont{T.}~\bibnamefont{Siro}},
  \bibinfo{author}{\bibfnamefont{L.}~\bibnamefont{Liang}},
  \bibinfo{author}{\bibfnamefont{M.}~\bibnamefont{Troyer}},
  \bibinfo{author}{\bibfnamefont{A.}~\bibnamefont{Harju}}, \bibnamefont{and}
  \bibinfo{author}{\bibfnamefont{P.}~\bibnamefont{T\"orm\"a}},
  \bibinfo{journal}{Phys. Rev. Lett.} \textbf{\bibinfo{volume}{116}},
  \bibinfo{pages}{225305} (\bibinfo{year}{2016}),
  \urlprefix\url{http://link.aps.org/doi/10.1103/PhysRevLett.116.225305}.

\bibitem[{\citenamefont{Peano et~al.}(2016)\citenamefont{Peano, Houde, Brendel,
  Marquardt, and Clerk}}]{peano2015topological}
\bibinfo{author}{\bibfnamefont{V.}~\bibnamefont{Peano}},
  \bibinfo{author}{\bibfnamefont{M.}~\bibnamefont{Houde}},
  \bibinfo{author}{\bibfnamefont{C.}~\bibnamefont{Brendel}},
  \bibinfo{author}{\bibfnamefont{F.}~\bibnamefont{Marquardt}},
  \bibnamefont{and} \bibinfo{author}{\bibfnamefont{A.~A.} \bibnamefont{Clerk}},
  \bibinfo{journal}{Nature communications} \textbf{\bibinfo{volume}{7}},
  \bibinfo{pages}{10779} (\bibinfo{year}{2016}),
  \urlprefix\url{http://dx.doi.org/10.1038/ncomms10779}.

\bibitem[{\citenamefont{Wang et~al.}(2009)\citenamefont{Wang, Chong,
  Joannopoulos, and Soljacic}}]{Soljacic2009}
\bibinfo{author}{\bibfnamefont{Z.}~\bibnamefont{Wang}},
  \bibinfo{author}{\bibfnamefont{Y.}~\bibnamefont{Chong}},
  \bibinfo{author}{\bibfnamefont{J.~D.} \bibnamefont{Joannopoulos}},
  \bibnamefont{and} \bibinfo{author}{\bibfnamefont{M.}~\bibnamefont{Soljacic}},
  \bibinfo{journal}{Nature} \textbf{\bibinfo{volume}{461}},
  \bibinfo{pages}{772775} (\bibinfo{year}{2009}), ISSN
  \bibinfo{issn}{0028-0836}.

\bibitem[{\citenamefont{Lu et~al.}(2014)\citenamefont{Lu, Joannopoulos, and
  Soljacic}}]{Soljacic2014}
\bibinfo{author}{\bibfnamefont{L.}~\bibnamefont{Lu}},
  \bibinfo{author}{\bibfnamefont{J.~D.} \bibnamefont{Joannopoulos}},
  \bibnamefont{and} \bibinfo{author}{\bibfnamefont{M.}~\bibnamefont{Soljacic}},
  \bibinfo{journal}{Nature Photonics} \textbf{\bibinfo{volume}{8}},
  \bibinfo{pages}{821} (\bibinfo{year}{2014}).

\bibitem[{\citenamefont{Aidelsburger et~al.}(2015)\citenamefont{Aidelsburger,
  Lohse, Schweizer, Atala, Barreiro, Nascimbene, Cooper, Bloch, and
  Goldman}}]{aidelsburger2015measuring}
\bibinfo{author}{\bibfnamefont{M.}~\bibnamefont{Aidelsburger}},
  \bibinfo{author}{\bibfnamefont{M.}~\bibnamefont{Lohse}},
  \bibinfo{author}{\bibfnamefont{C.}~\bibnamefont{Schweizer}},
  \bibinfo{author}{\bibfnamefont{M.}~\bibnamefont{Atala}},
  \bibinfo{author}{\bibfnamefont{J.~T.} \bibnamefont{Barreiro}},
  \bibinfo{author}{\bibfnamefont{S.}~\bibnamefont{Nascimbene}},
  \bibinfo{author}{\bibfnamefont{N.}~\bibnamefont{Cooper}},
  \bibinfo{author}{\bibfnamefont{I.}~\bibnamefont{Bloch}}, \bibnamefont{and}
  \bibinfo{author}{\bibfnamefont{N.}~\bibnamefont{Goldman}},
  \bibinfo{journal}{Nature Physics} \textbf{\bibinfo{volume}{11}},
  \bibinfo{pages}{162} (\bibinfo{year}{2015}).

\bibitem[{\citenamefont{Skirlo et~al.}(2015)\citenamefont{Skirlo, Lu, Igarashi,
  Yan, Joannopoulos, and Solja\ifmmode \check{c}\else
  \v{c}\fi{}i\ifmmode~\acute{c}\else \'{c}\fi{}}}]{PhysRevLett.115.253901}
\bibinfo{author}{\bibfnamefont{S.~A.} \bibnamefont{Skirlo}},
  \bibinfo{author}{\bibfnamefont{L.}~\bibnamefont{Lu}},
  \bibinfo{author}{\bibfnamefont{Y.}~\bibnamefont{Igarashi}},
  \bibinfo{author}{\bibfnamefont{Q.}~\bibnamefont{Yan}},
  \bibinfo{author}{\bibfnamefont{J.}~\bibnamefont{Joannopoulos}},
  \bibnamefont{and}
  \bibinfo{author}{\bibfnamefont{M.}~\bibnamefont{Solja\ifmmode \check{c}\else
  \v{c}\fi{}i\ifmmode~\acute{c}\else \'{c}\fi{}}}, \bibinfo{journal}{Phys. Rev.
  Lett.} \textbf{\bibinfo{volume}{115}}, \bibinfo{pages}{253901}
  (\bibinfo{year}{2015}),
  \urlprefix\url{http://link.aps.org/doi/10.1103/PhysRevLett.115.253901}.

\bibitem[{\citenamefont{Umucal\ifmmode \imath \else~\i \fi{}lar and
  Carusotto}(2012)}]{PhysRevLett.108.206809}
\bibinfo{author}{\bibfnamefont{R.~O.} \bibnamefont{Umucal\ifmmode \imath
  \else~\i \fi{}lar}} \bibnamefont{and}
  \bibinfo{author}{\bibfnamefont{I.}~\bibnamefont{Carusotto}},
  \bibinfo{journal}{Phys. Rev. Lett.} \textbf{\bibinfo{volume}{108}},
  \bibinfo{pages}{206809} (\bibinfo{year}{2012}),
  \urlprefix\url{http://link.aps.org/doi/10.1103/PhysRevLett.108.206809}.

\bibitem[{\citenamefont{Wang et~al.}(2014)\citenamefont{Wang, Potter, and
  Senthil}}]{wang2014classification}
\bibinfo{author}{\bibfnamefont{C.}~\bibnamefont{Wang}},
  \bibinfo{author}{\bibfnamefont{A.~C.} \bibnamefont{Potter}},
  \bibnamefont{and} \bibinfo{author}{\bibfnamefont{T.}~\bibnamefont{Senthil}},
  \bibinfo{journal}{Science} \textbf{\bibinfo{volume}{343}},
  \bibinfo{pages}{629} (\bibinfo{year}{2014}).

\bibitem[{\citenamefont{Peotta and
  T{\"o}rm{\"a}}(2015)}]{peotta2015superfluidity}
\bibinfo{author}{\bibfnamefont{S.}~\bibnamefont{Peotta}} \bibnamefont{and}
  \bibinfo{author}{\bibfnamefont{P.}~\bibnamefont{T{\"o}rm{\"a}}},
  \bibinfo{journal}{Nature communications} \textbf{\bibinfo{volume}{6}},
  \bibinfo{pages}{8944} (\bibinfo{year}{2015}).

\bibitem[{\citenamefont{Nalitov
  et~al.}(2015{\natexlab{a}})\citenamefont{Nalitov, Solnyshkov, and
  Malpuech}}]{nalitov2015polariton}
\bibinfo{author}{\bibfnamefont{A.}~\bibnamefont{Nalitov}},
  \bibinfo{author}{\bibfnamefont{D.}~\bibnamefont{Solnyshkov}},
  \bibnamefont{and} \bibinfo{author}{\bibfnamefont{G.}~\bibnamefont{Malpuech}},
  \bibinfo{journal}{Physical review letters} \textbf{\bibinfo{volume}{114}},
  \bibinfo{pages}{116401} (\bibinfo{year}{2015}{\natexlab{a}}).

\bibitem[{\citenamefont{Karzig et~al.}(2015)\citenamefont{Karzig, Bardyn,
  Lindner, and Refael}}]{KarzigPRX2015}
\bibinfo{author}{\bibfnamefont{T.}~\bibnamefont{Karzig}},
  \bibinfo{author}{\bibfnamefont{C.-E.} \bibnamefont{Bardyn}},
  \bibinfo{author}{\bibfnamefont{N.~H.} \bibnamefont{Lindner}},
  \bibnamefont{and} \bibinfo{author}{\bibfnamefont{G.}~\bibnamefont{Refael}},
  \bibinfo{journal}{Phys. Rev. X} \textbf{\bibinfo{volume}{5}},
  \bibinfo{pages}{031001} (\bibinfo{year}{2015}),
  \urlprefix\url{http://link.aps.org/doi/10.1103/PhysRevX.5.031001}.

\bibitem[{\citenamefont{Bardyn et~al.}(2015)\citenamefont{Bardyn, Karzig,
  Refael, and Liew}}]{LiewPRB2015}
\bibinfo{author}{\bibfnamefont{C.-E.} \bibnamefont{Bardyn}},
  \bibinfo{author}{\bibfnamefont{T.}~\bibnamefont{Karzig}},
  \bibinfo{author}{\bibfnamefont{G.}~\bibnamefont{Refael}}, \bibnamefont{and}
  \bibinfo{author}{\bibfnamefont{T.~C.~H.} \bibnamefont{Liew}},
  \bibinfo{journal}{Phys. Rev. B} \textbf{\bibinfo{volume}{91}},
  \bibinfo{pages}{161413} (\bibinfo{year}{2015}),
  \urlprefix\url{http://link.aps.org/doi/10.1103/PhysRevB.91.161413}.

\bibitem[{\citenamefont{Yi and Karzig}(2016)}]{PhysRevB.93.104303}
\bibinfo{author}{\bibfnamefont{K.}~\bibnamefont{Yi}} \bibnamefont{and}
  \bibinfo{author}{\bibfnamefont{T.}~\bibnamefont{Karzig}},
  \bibinfo{journal}{Phys. Rev. B} \textbf{\bibinfo{volume}{93}},
  \bibinfo{pages}{104303} (\bibinfo{year}{2016}),
  \urlprefix\url{http://link.aps.org/doi/10.1103/PhysRevB.93.104303}.

\bibitem[{\citenamefont{Bardyn et~al.}(2016)\citenamefont{Bardyn, Karzig,
  Refael, and Liew}}]{PhysRevB.93.020502}
\bibinfo{author}{\bibfnamefont{C.-E.} \bibnamefont{Bardyn}},
  \bibinfo{author}{\bibfnamefont{T.}~\bibnamefont{Karzig}},
  \bibinfo{author}{\bibfnamefont{G.}~\bibnamefont{Refael}}, \bibnamefont{and}
  \bibinfo{author}{\bibfnamefont{T.~C.~H.} \bibnamefont{Liew}},
  \bibinfo{journal}{Phys. Rev. B} \textbf{\bibinfo{volume}{93}},
  \bibinfo{pages}{020502} (\bibinfo{year}{2016}),
  \urlprefix\url{http://link.aps.org/doi/10.1103/PhysRevB.93.020502}.

\bibitem[{\citenamefont{Bleu et~al.}(2016)\citenamefont{Bleu, Solnyshkov, and
  Malpuech}}]{PhysRevB.93.085438}
\bibinfo{author}{\bibfnamefont{O.}~\bibnamefont{Bleu}},
  \bibinfo{author}{\bibfnamefont{D.~D.} \bibnamefont{Solnyshkov}},
  \bibnamefont{and} \bibinfo{author}{\bibfnamefont{G.}~\bibnamefont{Malpuech}},
  \bibinfo{journal}{Phys. Rev. B} \textbf{\bibinfo{volume}{93}},
  \bibinfo{pages}{085438} (\bibinfo{year}{2016}),
  \urlprefix\url{http://link.aps.org/doi/10.1103/PhysRevB.93.085438}.

\bibitem[{\citenamefont{Carusotto and Ciuti}(2013)}]{carusotto2013quantum}
\bibinfo{author}{\bibfnamefont{I.}~\bibnamefont{Carusotto}} \bibnamefont{and}
  \bibinfo{author}{\bibfnamefont{C.}~\bibnamefont{Ciuti}},
  \bibinfo{journal}{Reviews of Modern Physics} \textbf{\bibinfo{volume}{85}},
  \bibinfo{pages}{299} (\bibinfo{year}{2013}).

\bibitem[{\citenamefont{Shelykh
  et~al.}(2009{\natexlab{a}})\citenamefont{Shelykh, Kavokin, Rubo, Liew, and
  Malpuech}}]{shelykh2009polariton}
\bibinfo{author}{\bibfnamefont{I.}~\bibnamefont{Shelykh}},
  \bibinfo{author}{\bibfnamefont{A.}~\bibnamefont{Kavokin}},
  \bibinfo{author}{\bibfnamefont{Y.~G.} \bibnamefont{Rubo}},
  \bibinfo{author}{\bibfnamefont{T.}~\bibnamefont{Liew}}, \bibnamefont{and}
  \bibinfo{author}{\bibfnamefont{G.}~\bibnamefont{Malpuech}},
  \bibinfo{journal}{Semiconductor Science and Technology}
  \textbf{\bibinfo{volume}{25}}, \bibinfo{pages}{013001}
  (\bibinfo{year}{2009}{\natexlab{a}}).

\bibitem[{\citenamefont{Lin et~al.}(2009)\citenamefont{Lin, Compton,
  Jimenez-Garcia, Porto, and Spielman}}]{lin2009synthetic}
\bibinfo{author}{\bibfnamefont{Y.-J.} \bibnamefont{Lin}},
  \bibinfo{author}{\bibfnamefont{R.~L.} \bibnamefont{Compton}},
  \bibinfo{author}{\bibfnamefont{K.}~\bibnamefont{Jimenez-Garcia}},
  \bibinfo{author}{\bibfnamefont{J.~V.} \bibnamefont{Porto}}, \bibnamefont{and}
  \bibinfo{author}{\bibfnamefont{I.~B.} \bibnamefont{Spielman}},
  \bibinfo{journal}{Nature} \textbf{\bibinfo{volume}{462}},
  \bibinfo{pages}{628} (\bibinfo{year}{2009}).

\bibitem[{\citenamefont{Lin et~al.}(2011)\citenamefont{Lin, Jimenez-Garcia, and
  Spielman}}]{lin2011spin}
\bibinfo{author}{\bibfnamefont{Y.-J.} \bibnamefont{Lin}},
  \bibinfo{author}{\bibfnamefont{K.}~\bibnamefont{Jimenez-Garcia}},
  \bibnamefont{and} \bibinfo{author}{\bibfnamefont{I.}~\bibnamefont{Spielman}},
  \bibinfo{journal}{Nature} \textbf{\bibinfo{volume}{471}}, \bibinfo{pages}{83}
  (\bibinfo{year}{2011}).

\bibitem[{\citenamefont{Rechtsman et~al.}(2013)\citenamefont{Rechtsman, Zeuner,
  Plotnik, Lumer, Podolsky, Dreisow, Nolte, Segev, and
  Szameit}}]{rechtsman2013photonic}
\bibinfo{author}{\bibfnamefont{M.~C.} \bibnamefont{Rechtsman}},
  \bibinfo{author}{\bibfnamefont{J.~M.} \bibnamefont{Zeuner}},
  \bibinfo{author}{\bibfnamefont{Y.}~\bibnamefont{Plotnik}},
  \bibinfo{author}{\bibfnamefont{Y.}~\bibnamefont{Lumer}},
  \bibinfo{author}{\bibfnamefont{D.}~\bibnamefont{Podolsky}},
  \bibinfo{author}{\bibfnamefont{F.}~\bibnamefont{Dreisow}},
  \bibinfo{author}{\bibfnamefont{S.}~\bibnamefont{Nolte}},
  \bibinfo{author}{\bibfnamefont{M.}~\bibnamefont{Segev}}, \bibnamefont{and}
  \bibinfo{author}{\bibfnamefont{A.}~\bibnamefont{Szameit}},
  \bibinfo{journal}{Nature} \textbf{\bibinfo{volume}{496}},
  \bibinfo{pages}{196} (\bibinfo{year}{2013}).

\bibitem[{\citenamefont{Hafezi et~al.}(2011)\citenamefont{Hafezi, Demler,
  Lukin, and Taylor}}]{hafezi2011robust}
\bibinfo{author}{\bibfnamefont{M.}~\bibnamefont{Hafezi}},
  \bibinfo{author}{\bibfnamefont{E.~A.} \bibnamefont{Demler}},
  \bibinfo{author}{\bibfnamefont{M.~D.} \bibnamefont{Lukin}}, \bibnamefont{and}
  \bibinfo{author}{\bibfnamefont{J.~M.} \bibnamefont{Taylor}},
  \bibinfo{journal}{Nature Physics} \textbf{\bibinfo{volume}{7}},
  \bibinfo{pages}{907} (\bibinfo{year}{2011}).

\bibitem[{\citenamefont{Hafezi et~al.}(2013)\citenamefont{Hafezi, Mittal, Fan,
  Migdall, and Taylor}}]{hafezi2013imaging}
\bibinfo{author}{\bibfnamefont{M.}~\bibnamefont{Hafezi}},
  \bibinfo{author}{\bibfnamefont{S.}~\bibnamefont{Mittal}},
  \bibinfo{author}{\bibfnamefont{J.}~\bibnamefont{Fan}},
  \bibinfo{author}{\bibfnamefont{A.}~\bibnamefont{Migdall}}, \bibnamefont{and}
  \bibinfo{author}{\bibfnamefont{J.}~\bibnamefont{Taylor}},
  \bibinfo{journal}{Nature Photonics} \textbf{\bibinfo{volume}{7}},
  \bibinfo{pages}{1001} (\bibinfo{year}{2013}).

\bibitem[{\citenamefont{Khanikaev et~al.}(2013)\citenamefont{Khanikaev,
  Mousavi, Tse, Kargarian, MacDonald, and Shvets}}]{khanikaev2013photonic}
\bibinfo{author}{\bibfnamefont{A.~B.} \bibnamefont{Khanikaev}},
  \bibinfo{author}{\bibfnamefont{S.~H.} \bibnamefont{Mousavi}},
  \bibinfo{author}{\bibfnamefont{W.-K.} \bibnamefont{Tse}},
  \bibinfo{author}{\bibfnamefont{M.}~\bibnamefont{Kargarian}},
  \bibinfo{author}{\bibfnamefont{A.~H.} \bibnamefont{MacDonald}},
  \bibnamefont{and} \bibinfo{author}{\bibfnamefont{G.}~\bibnamefont{Shvets}},
  \bibinfo{journal}{Nature materials} \textbf{\bibinfo{volume}{15}},
  \bibinfo{pages}{542–548} (\bibinfo{year}{2013}).

\bibitem[{\citenamefont{Haldane and Raghu}(2008)}]{haldane2008possible}
\bibinfo{author}{\bibfnamefont{F.}~\bibnamefont{Haldane}} \bibnamefont{and}
  \bibinfo{author}{\bibfnamefont{S.}~\bibnamefont{Raghu}},
  \bibinfo{journal}{Physical review letters} \textbf{\bibinfo{volume}{100}},
  \bibinfo{pages}{013904} (\bibinfo{year}{2008}).

\bibitem[{\citenamefont{Nalitov
  et~al.}(2015{\natexlab{b}})\citenamefont{Nalitov, Malpuech,
  Ter\ifmmode~\mbox{\c{c}}\else \c{c}\fi{}as, and
  Solnyshkov}}]{PhysRevLett.114.026803}
\bibinfo{author}{\bibfnamefont{A.~V.} \bibnamefont{Nalitov}},
  \bibinfo{author}{\bibfnamefont{G.}~\bibnamefont{Malpuech}},
  \bibinfo{author}{\bibfnamefont{H.}~\bibnamefont{Ter\ifmmode~\mbox{\c{c}}\else
  \c{c}\fi{}as}}, \bibnamefont{and} \bibinfo{author}{\bibfnamefont{D.~D.}
  \bibnamefont{Solnyshkov}}, \bibinfo{journal}{Phys. Rev. Lett.}
  \textbf{\bibinfo{volume}{114}}, \bibinfo{pages}{026803}
  (\bibinfo{year}{2015}{\natexlab{b}}),
  \urlprefix\url{http://link.aps.org/doi/10.1103/PhysRevLett.114.026803}.

\bibitem[{\citenamefont{Shelykh
  et~al.}(2009{\natexlab{b}})\citenamefont{Shelykh, Pavlovic, Solnyshkov, and
  Malpuech}}]{PhysRevLett.102.046407}
\bibinfo{author}{\bibfnamefont{I.~A.} \bibnamefont{Shelykh}},
  \bibinfo{author}{\bibfnamefont{G.}~\bibnamefont{Pavlovic}},
  \bibinfo{author}{\bibfnamefont{D.~D.} \bibnamefont{Solnyshkov}},
  \bibnamefont{and} \bibinfo{author}{\bibfnamefont{G.}~\bibnamefont{Malpuech}},
  \bibinfo{journal}{Phys. Rev. Lett.} \textbf{\bibinfo{volume}{102}},
  \bibinfo{pages}{046407} (\bibinfo{year}{2009}{\natexlab{b}}),
  \urlprefix\url{http://link.aps.org/doi/10.1103/PhysRevLett.102.046407}.

\bibitem[{\citenamefont{Fischer et~al.}(2014)\citenamefont{Fischer, Brodbeck,
  Chernenko, Lederer, Rahimi-Iman, Amthor, Kulakovskii, Worschech, Kamp, Durnev
  et~al.}}]{PhysRevLett.112.093902}
\bibinfo{author}{\bibfnamefont{J.}~\bibnamefont{Fischer}},
  \bibinfo{author}{\bibfnamefont{S.}~\bibnamefont{Brodbeck}},
  \bibinfo{author}{\bibfnamefont{A.~V.} \bibnamefont{Chernenko}},
  \bibinfo{author}{\bibfnamefont{I.}~\bibnamefont{Lederer}},
  \bibinfo{author}{\bibfnamefont{A.}~\bibnamefont{Rahimi-Iman}},
  \bibinfo{author}{\bibfnamefont{M.}~\bibnamefont{Amthor}},
  \bibinfo{author}{\bibfnamefont{V.~D.} \bibnamefont{Kulakovskii}},
  \bibinfo{author}{\bibfnamefont{L.}~\bibnamefont{Worschech}},
  \bibinfo{author}{\bibfnamefont{M.}~\bibnamefont{Kamp}},
  \bibinfo{author}{\bibfnamefont{M.}~\bibnamefont{Durnev}},
  \bibnamefont{et~al.}, \bibinfo{journal}{Phys. Rev. Lett.}
  \textbf{\bibinfo{volume}{112}}, \bibinfo{pages}{093902}
  (\bibinfo{year}{2014}),
  \urlprefix\url{http://link.aps.org/doi/10.1103/PhysRevLett.112.093902}.

\bibitem[{\citenamefont{Sturm et~al.}(2015)\citenamefont{Sturm, Solnyshkov,
  Krebs, Lema\^{\i}tre, Sagnes, Galopin, Amo, Malpuech, and
  Bloch}}]{PhysRevB.91.155130}
\bibinfo{author}{\bibfnamefont{C.}~\bibnamefont{Sturm}},
  \bibinfo{author}{\bibfnamefont{D.}~\bibnamefont{Solnyshkov}},
  \bibinfo{author}{\bibfnamefont{O.}~\bibnamefont{Krebs}},
  \bibinfo{author}{\bibfnamefont{A.}~\bibnamefont{Lema\^{\i}tre}},
  \bibinfo{author}{\bibfnamefont{I.}~\bibnamefont{Sagnes}},
  \bibinfo{author}{\bibfnamefont{E.}~\bibnamefont{Galopin}},
  \bibinfo{author}{\bibfnamefont{A.}~\bibnamefont{Amo}},
  \bibinfo{author}{\bibfnamefont{G.}~\bibnamefont{Malpuech}}, \bibnamefont{and}
  \bibinfo{author}{\bibfnamefont{J.}~\bibnamefont{Bloch}},
  \bibinfo{journal}{Phys. Rev. B} \textbf{\bibinfo{volume}{91}},
  \bibinfo{pages}{155130} (\bibinfo{year}{2015}),
  \urlprefix\url{http://link.aps.org/doi/10.1103/PhysRevB.91.155130}.

\bibitem[{\citenamefont{Kavokin et~al.}(2005)\citenamefont{Kavokin, Malpuech,
  and Glazov}}]{PhysRevLett.95.136601}
\bibinfo{author}{\bibfnamefont{A.}~\bibnamefont{Kavokin}},
  \bibinfo{author}{\bibfnamefont{G.}~\bibnamefont{Malpuech}}, \bibnamefont{and}
  \bibinfo{author}{\bibfnamefont{M.}~\bibnamefont{Glazov}},
  \bibinfo{journal}{Phys. Rev. Lett.} \textbf{\bibinfo{volume}{95}},
  \bibinfo{pages}{136601} (\bibinfo{year}{2005}),
  \urlprefix\url{http://link.aps.org/doi/10.1103/PhysRevLett.95.136601}.

\bibitem[{\citenamefont{Leyder et~al.}(2007)\citenamefont{Leyder, Romanelli,
  Karr, Giacobino, Liew, Glazov, Kavokin, Malpuech, and
  Bramati}}]{leyder2007observation}
\bibinfo{author}{\bibfnamefont{C.}~\bibnamefont{Leyder}},
  \bibinfo{author}{\bibfnamefont{M.}~\bibnamefont{Romanelli}},
  \bibinfo{author}{\bibfnamefont{J.~P.} \bibnamefont{Karr}},
  \bibinfo{author}{\bibfnamefont{E.}~\bibnamefont{Giacobino}},
  \bibinfo{author}{\bibfnamefont{T.~C.} \bibnamefont{Liew}},
  \bibinfo{author}{\bibfnamefont{M.~M.} \bibnamefont{Glazov}},
  \bibinfo{author}{\bibfnamefont{A.~V.} \bibnamefont{Kavokin}},
  \bibinfo{author}{\bibfnamefont{G.}~\bibnamefont{Malpuech}}, \bibnamefont{and}
  \bibinfo{author}{\bibfnamefont{A.}~\bibnamefont{Bramati}},
  \bibinfo{journal}{Nature Physics} \textbf{\bibinfo{volume}{3}},
  \bibinfo{pages}{628} (\bibinfo{year}{2007}).

\bibitem[{\citenamefont{Sala et~al.}(2015)\citenamefont{Sala, Solnyshkov,
  Carusotto, Jacqmin, Lema\^{\i}tre, Ter\ifmmode~\mbox{\c{c}}\else
  \c{c}\fi{}as, Nalitov, Abbarchi, Galopin, Sagnes et~al.}}]{PhysRevX.5.011034}
\bibinfo{author}{\bibfnamefont{V.~G.} \bibnamefont{Sala}},
  \bibinfo{author}{\bibfnamefont{D.~D.} \bibnamefont{Solnyshkov}},
  \bibinfo{author}{\bibfnamefont{I.}~\bibnamefont{Carusotto}},
  \bibinfo{author}{\bibfnamefont{T.}~\bibnamefont{Jacqmin}},
  \bibinfo{author}{\bibfnamefont{A.}~\bibnamefont{Lema\^{\i}tre}},
  \bibinfo{author}{\bibfnamefont{H.}~\bibnamefont{Ter\ifmmode~\mbox{\c{c}}\else
  \c{c}\fi{}as}}, \bibinfo{author}{\bibfnamefont{A.}~\bibnamefont{Nalitov}},
  \bibinfo{author}{\bibfnamefont{M.}~\bibnamefont{Abbarchi}},
  \bibinfo{author}{\bibfnamefont{E.}~\bibnamefont{Galopin}},
  \bibinfo{author}{\bibfnamefont{I.}~\bibnamefont{Sagnes}},
  \bibnamefont{et~al.}, \bibinfo{journal}{Phys. Rev. X}
  \textbf{\bibinfo{volume}{5}}, \bibinfo{pages}{011034} (\bibinfo{year}{2015}),
  \urlprefix\url{http://link.aps.org/doi/10.1103/PhysRevX.5.011034}.

\bibitem[{\citenamefont{Carusotto and Ciuti}(2004)}]{PhysRevLett.93.166401}
\bibinfo{author}{\bibfnamefont{I.}~\bibnamefont{Carusotto}} \bibnamefont{and}
  \bibinfo{author}{\bibfnamefont{C.}~\bibnamefont{Ciuti}},
  \bibinfo{journal}{Phys. Rev. Lett.} \textbf{\bibinfo{volume}{93}},
  \bibinfo{pages}{166401} (\bibinfo{year}{2004}),
  \urlprefix\url{http://link.aps.org/doi/10.1103/PhysRevLett.93.166401}.

\bibitem[{\citenamefont{Hivet et~al.}(2012)\citenamefont{Hivet, Flayac,
  Solnyshkov, Tanese, Boulier, Andreoli, Giacobino, Bloch, Bramati, Malpuech
  et~al.}}]{hivet2012half}
\bibinfo{author}{\bibfnamefont{R.}~\bibnamefont{Hivet}},
  \bibinfo{author}{\bibfnamefont{H.}~\bibnamefont{Flayac}},
  \bibinfo{author}{\bibfnamefont{D.}~\bibnamefont{Solnyshkov}},
  \bibinfo{author}{\bibfnamefont{D.}~\bibnamefont{Tanese}},
  \bibinfo{author}{\bibfnamefont{T.}~\bibnamefont{Boulier}},
  \bibinfo{author}{\bibfnamefont{D.}~\bibnamefont{Andreoli}},
  \bibinfo{author}{\bibfnamefont{E.}~\bibnamefont{Giacobino}},
  \bibinfo{author}{\bibfnamefont{J.}~\bibnamefont{Bloch}},
  \bibinfo{author}{\bibfnamefont{A.}~\bibnamefont{Bramati}},
  \bibinfo{author}{\bibfnamefont{G.}~\bibnamefont{Malpuech}},
  \bibnamefont{et~al.}, \bibinfo{journal}{Nature Physics}
  \textbf{\bibinfo{volume}{8}}, \bibinfo{pages}{724} (\bibinfo{year}{2012}).

\bibitem[{\citenamefont{Dominici et~al.}(2014)\citenamefont{Dominici,
  Dagvadorj, Fellows, Donati, Ballarini, De~Giorgi, Marchetti, Piccirillo,
  Marrucci, Bramati et~al.}}]{dominici2014vortex}
\bibinfo{author}{\bibfnamefont{L.}~\bibnamefont{Dominici}},
  \bibinfo{author}{\bibfnamefont{G.}~\bibnamefont{Dagvadorj}},
  \bibinfo{author}{\bibfnamefont{J.}~\bibnamefont{Fellows}},
  \bibinfo{author}{\bibfnamefont{S.}~\bibnamefont{Donati}},
  \bibinfo{author}{\bibfnamefont{D.}~\bibnamefont{Ballarini}},
  \bibinfo{author}{\bibfnamefont{M.}~\bibnamefont{De~Giorgi}},
  \bibinfo{author}{\bibfnamefont{F.}~\bibnamefont{Marchetti}},
  \bibinfo{author}{\bibfnamefont{B.}~\bibnamefont{Piccirillo}},
  \bibinfo{author}{\bibfnamefont{L.}~\bibnamefont{Marrucci}},
  \bibinfo{author}{\bibfnamefont{A.}~\bibnamefont{Bramati}},
  \bibnamefont{et~al.}, \bibinfo{journal}{arXiv preprint arXiv:1403.0487}
  (\bibinfo{year}{2014}).

\bibitem[{\citenamefont{Jacqmin et~al.}(2014)\citenamefont{Jacqmin, Carusotto,
  Sagnes, Abbarchi, Solnyshkov, Malpuech, Galopin, Lema\^{\i}tre, Bloch, and
  Amo}}]{PhysRevLett.112.116402}
\bibinfo{author}{\bibfnamefont{T.}~\bibnamefont{Jacqmin}},
  \bibinfo{author}{\bibfnamefont{I.}~\bibnamefont{Carusotto}},
  \bibinfo{author}{\bibfnamefont{I.}~\bibnamefont{Sagnes}},
  \bibinfo{author}{\bibfnamefont{M.}~\bibnamefont{Abbarchi}},
  \bibinfo{author}{\bibfnamefont{D.~D.} \bibnamefont{Solnyshkov}},
  \bibinfo{author}{\bibfnamefont{G.}~\bibnamefont{Malpuech}},
  \bibinfo{author}{\bibfnamefont{E.}~\bibnamefont{Galopin}},
  \bibinfo{author}{\bibfnamefont{A.}~\bibnamefont{Lema\^{\i}tre}},
  \bibinfo{author}{\bibfnamefont{J.}~\bibnamefont{Bloch}}, \bibnamefont{and}
  \bibinfo{author}{\bibfnamefont{A.}~\bibnamefont{Amo}},
  \bibinfo{journal}{Phys. Rev. Lett.} \textbf{\bibinfo{volume}{112}},
  \bibinfo{pages}{116402} (\bibinfo{year}{2014}),
  \urlprefix\url{http://link.aps.org/doi/10.1103/PhysRevLett.112.116402}.

\bibitem[{\citenamefont{Zhang et~al.}(2015)\citenamefont{Zhang, Zhao, Yao, and
  Yang}}]{zhang2015quantum}
\bibinfo{author}{\bibfnamefont{J.}~\bibnamefont{Zhang}},
  \bibinfo{author}{\bibfnamefont{B.}~\bibnamefont{Zhao}},
  \bibinfo{author}{\bibfnamefont{Y.}~\bibnamefont{Yao}}, \bibnamefont{and}
  \bibinfo{author}{\bibfnamefont{Z.}~\bibnamefont{Yang}},
  \bibinfo{journal}{Scientific reports} \textbf{\bibinfo{volume}{5}},
  \bibinfo{pages}{10629} (\bibinfo{year}{2015}).

\bibitem[{\citenamefont{Cheng et~al.}(2016)\citenamefont{Cheng, Jouvaud, Ni,
  Mousavi, Genack, and Khanikaev}}]{cheng2016robust}
\bibinfo{author}{\bibfnamefont{X.}~\bibnamefont{Cheng}},
  \bibinfo{author}{\bibfnamefont{C.}~\bibnamefont{Jouvaud}},
  \bibinfo{author}{\bibfnamefont{X.}~\bibnamefont{Ni}},
  \bibinfo{author}{\bibfnamefont{S.~H.} \bibnamefont{Mousavi}},
  \bibinfo{author}{\bibfnamefont{A.~Z.} \bibnamefont{Genack}},
  \bibnamefont{and} \bibinfo{author}{\bibfnamefont{A.~B.}
  \bibnamefont{Khanikaev}}, \bibinfo{journal}{Nature materials}
  \textbf{\bibinfo{volume}{12}}, \bibinfo{pages}{233} (\bibinfo{year}{2016}).

\bibitem[{\citenamefont{Solnyshkov et~al.}(2008)\citenamefont{Solnyshkov,
  Shelykh, Gippius, Kavokin, and Malpuech}}]{PhysRevB.77.045314}
\bibinfo{author}{\bibfnamefont{D.~D.} \bibnamefont{Solnyshkov}},
  \bibinfo{author}{\bibfnamefont{I.~A.} \bibnamefont{Shelykh}},
  \bibinfo{author}{\bibfnamefont{N.~A.} \bibnamefont{Gippius}},
  \bibinfo{author}{\bibfnamefont{A.~V.} \bibnamefont{Kavokin}},
  \bibnamefont{and} \bibinfo{author}{\bibfnamefont{G.}~\bibnamefont{Malpuech}},
  \bibinfo{journal}{Phys. Rev. B} \textbf{\bibinfo{volume}{77}},
  \bibinfo{pages}{045314} (\bibinfo{year}{2008}),
  \urlprefix\url{http://link.aps.org/doi/10.1103/PhysRevB.77.045314}.

\bibitem[{\citenamefont{Renucci et~al.}(2005)\citenamefont{Renucci, Amand,
  Marie, Senellart, Bloch, Sermage, and Kavokin}}]{Renucci2005}
\bibinfo{author}{\bibfnamefont{P.}~\bibnamefont{Renucci}},
  \bibinfo{author}{\bibfnamefont{T.}~\bibnamefont{Amand}},
  \bibinfo{author}{\bibfnamefont{X.}~\bibnamefont{Marie}},
  \bibinfo{author}{\bibfnamefont{P.}~\bibnamefont{Senellart}},
  \bibinfo{author}{\bibfnamefont{J.}~\bibnamefont{Bloch}},
  \bibinfo{author}{\bibfnamefont{B.}~\bibnamefont{Sermage}}, \bibnamefont{and}
  \bibinfo{author}{\bibfnamefont{K.~V.} \bibnamefont{Kavokin}},
  \bibinfo{journal}{Phys. Rev. B} \textbf{\bibinfo{volume}{72}},
  \bibinfo{pages}{075317} (\bibinfo{year}{2005}),
  \urlprefix\url{http://link.aps.org/doi/10.1103/PhysRevB.72.075317}.

\bibitem[{\citenamefont{Vladimirova et~al.}(2010)\citenamefont{Vladimirova,
  Cronenberger, Scalbert, Kavokin, Miard, Lema\^{\i}tre, Bloch, Solnyshkov,
  Malpuech, and Kavokin}}]{Vladimirova2010}
\bibinfo{author}{\bibfnamefont{M.}~\bibnamefont{Vladimirova}},
  \bibinfo{author}{\bibfnamefont{S.}~\bibnamefont{Cronenberger}},
  \bibinfo{author}{\bibfnamefont{D.}~\bibnamefont{Scalbert}},
  \bibinfo{author}{\bibfnamefont{K.~V.} \bibnamefont{Kavokin}},
  \bibinfo{author}{\bibfnamefont{A.}~\bibnamefont{Miard}},
  \bibinfo{author}{\bibfnamefont{A.}~\bibnamefont{Lema\^{\i}tre}},
  \bibinfo{author}{\bibfnamefont{J.}~\bibnamefont{Bloch}},
  \bibinfo{author}{\bibfnamefont{D.}~\bibnamefont{Solnyshkov}},
  \bibinfo{author}{\bibfnamefont{G.}~\bibnamefont{Malpuech}}, \bibnamefont{and}
  \bibinfo{author}{\bibfnamefont{A.~V.} \bibnamefont{Kavokin}},
  \bibinfo{journal}{Phys. Rev. B} \textbf{\bibinfo{volume}{82}},
  \bibinfo{pages}{075301} (\bibinfo{year}{2010}),
  \urlprefix\url{http://link.aps.org/doi/10.1103/PhysRevB.82.075301}.

\bibitem[{sup()}]{suppl}
\bibinfo{note}{See Supplemental Material at [URL will be inserted by publisher]
  for more details on the calculations.}

\bibitem[{\citenamefont{Hatsugai}(1993)}]{PhysRevLett.71.3697}
\bibinfo{author}{\bibfnamefont{Y.}~\bibnamefont{Hatsugai}},
  \bibinfo{journal}{Phys. Rev. Lett.} \textbf{\bibinfo{volume}{71}},
  \bibinfo{pages}{3697} (\bibinfo{year}{1993}),
  \urlprefix\url{http://link.aps.org/doi/10.1103/PhysRevLett.71.3697}.

\bibitem[{\citenamefont{Zarea and Sandler}(2009)}]{PhysRevB.79.165442}
\bibinfo{author}{\bibfnamefont{M.}~\bibnamefont{Zarea}} \bibnamefont{and}
  \bibinfo{author}{\bibfnamefont{N.}~\bibnamefont{Sandler}},
  \bibinfo{journal}{Phys. Rev. B} \textbf{\bibinfo{volume}{79}},
  \bibinfo{pages}{165442} (\bibinfo{year}{2009}),
  \urlprefix\url{http://link.aps.org/doi/10.1103/PhysRevB.79.165442}.

\bibitem[{\citenamefont{Fukui et~al.}(2005)\citenamefont{Fukui, Hatsugai, and
  Suzuki}}]{fukui2005chern}
\bibinfo{author}{\bibfnamefont{T.}~\bibnamefont{Fukui}},
  \bibinfo{author}{\bibfnamefont{Y.}~\bibnamefont{Hatsugai}}, \bibnamefont{and}
  \bibinfo{author}{\bibfnamefont{H.}~\bibnamefont{Suzuki}},
  \bibinfo{journal}{Journal of the Physical Society of Japan}
  \textbf{\bibinfo{volume}{74}}, \bibinfo{pages}{1674} (\bibinfo{year}{2005}).

\end{thebibliography}
 
\section{Supplemental material} 
 In this supplemental material, we present analytical results concerning the phase diagram of polariton graphene with applied magnetic field (no optical pumping), extending the case studied in \cite{nalitov2015polariton}. We also provide more details on the calculations for the non-linear case with TE-TM and Rashba spin-orbit coupling.
 
 \subsection{Linear approach with an applied magnetic field}
In this section, we derive the complete topological phase diagram of the polariton  $\mathbb{Z}$ topological insulator in the linear regime, under applied magnetic field. This is a simpler case with respect to the main text, but it has the advantage of the existence of analytical solution for each phase transition, as shown below.

  We recall the linear Hamiltonian of the polariton Chern ($\mathbb{Z}$) insulator system and study the topological transitions in the linear regime.
 The tight-binding Hamiltonian can be written as in the main text: 
\begin{eqnarray}
H_k=\begin{pmatrix}
\Delta \sigma_Z&F_k \\
F_k^+& \Delta \sigma_Z
\end{pmatrix} 
,\quad F_k=-\begin{pmatrix}
f_kJ&f_k^+\delta J \\
f_k^- \delta J&f_kJ 
\end{pmatrix}
\end{eqnarray}
where $\Delta=|x|^2 g_x \mu_B \frac{B}{2}$ is the Zeeman splitting due to the applied external field, the complex coefficients $f_k$ and $f_k^{\pm}$ are defined by:
\begin{eqnarray}
f_k &=&\sum_{j=1}^3\exp{(-i\textbf{kd}_{\phi_j})},\\  f_k^{\pm}&=&\sum_{j=1}^3 \exp{(-i[\textbf{kd}_{\phi_j}\mp 2\phi_j])}\nonumber
\end{eqnarray}

Due to the presence of the sublattice (A/B) pseudospin and of the polarization spin (+/-) degree of freedom, this Hamiltonian is a 4-by-4 matrix. $J$ is the tunneling coefficient between nearest neighbour micro-pillars (A/B) and $\delta J$ is the spin-orbit coupling (+/-). 
 Without magnetic field ($\Delta=0$), the diagonalization of this Hamiltonian gives 4 branches of dispersion with degeneracies around $K$ and $K'$ points (Fig. 5(a),(b)). We can already distinguish two configurations: $\delta J/J>0.5$ and $\delta J/J<0.5$. In the first case, a single degeneracy appears at each K point, leading to the formation of the Dirac cones. When the SOC term is below the critical value of $0.5$, the TE-TM splitting leads to trigonal warping, with three additional Dirac cones appearing in the $K-M$ directions. When the magnetic field is switched on, a gap opens between the two upper and the two lower branches (that is, between the "conduction" and the "valence" bands), characterized by a Chern number $C=\pm 2$ depending on the direction of the external field. Here, we would like to stress the difference between branches and bands. An energy band (e.g. conduction band) is the ensemble of eigenstates belonging to the same continuum of available energy states (a segment of the energy axis). A band may contain several branches, which can exhibit different values of energy for the same wavevector (e.g. because of spin splitting - Rashba, Dresselhaus, TE-TM, or Zeeman), or they can be completely degenerate.
 
  Then, when either the strength of the magnetic field or the SOC is increased, the gap closes at the $M$ point, and then reopens with the Chern numbers $C=\mp1$. This topological transition arises from the degeneracy at the $M$ point and is described by the  following analytical formula:
 \begin{equation}
 \Delta_1=\sqrt{J^2-4\delta J^2}
 \end{equation}
 
When the magnetic field is increased even further, other topological transitions occur at their turn. One is obtained with the opening of two additional gaps between the two lower and two upper branches (in the middle of the "conduction" band and of the "valence" band, correspondingly). This transition arises when the energy of the second branch at the $\Gamma$ point (the minimum of the 2nd branch) is equal to the energy of the first (the lowest) one at the $K$ point (the maximum of the 1st branch), and thus the system of 2 bands (each containing 2 branches) is split into 4 bands (each containing a single branch). The corresponding formula is:

  \begin{equation}
 \Delta_2=\frac{3(J^2-\delta J^2)}{2J}
 \end{equation}
The transition to the the third topological phase occurs when the middle gap closes at the $\Gamma$ point for $ \Delta_3=3J$. Above this field, the middle gap is trivial but the two other bandgaps are still topological.

We compute the Chern numbers of each band for the different topological phases using the standard gauge-independent and stable technique of \cite{fukui2005chern}.
\begin{equation}
C_n=\frac{1}{2i\pi}\int_{BZ} \mathbf{F}_n(\mathbf{k}) d\textbf{k}
\label{chern}  
\end{equation}
where the Berry connection $\mathbf{A}$ and the associated Berry curvature $\mathbf{F}$ are given by:
\begin{eqnarray}
\mathbf{A}_n(\mathbf{k})&=&\bra{\textbf{u}_{\textbf{k},n}}\nabla_{\textbf{k}} \ket{\textbf{u}_{\textbf{k},n}}\\
\mathbf{F}_n(\mathbf{k})&=&\nabla_{\textbf{k}}\times\mathbf{A}_n
\end{eqnarray}
 A complete phase diagram is plotted in Fig. 5(c) as a function of the SOC and the magnetic field with the corresponding band Chern numbers for each phase.  
  \begin{figure}[tbp]
 \begin{center}
 \includegraphics[scale=0.55]{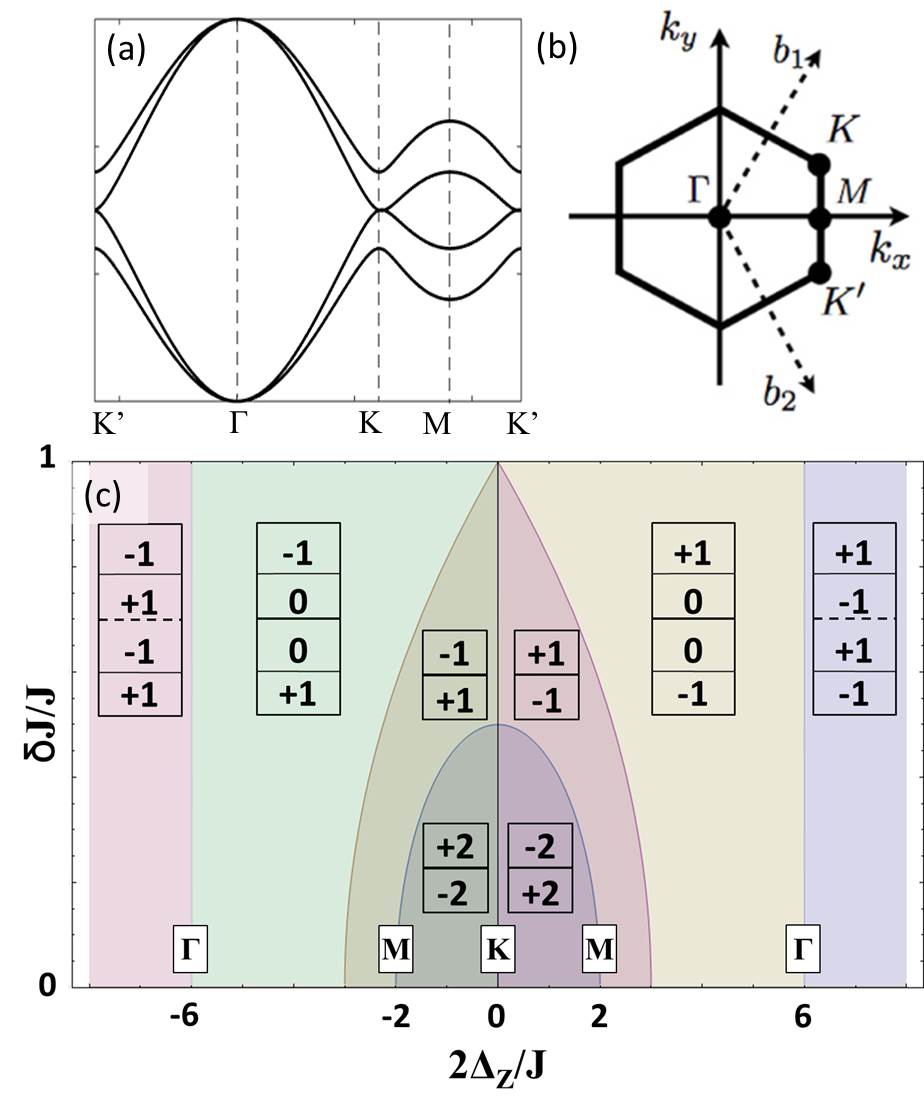}
 \caption{ (Color online) (a) Polaritonic graphene dispersion ($\delta J=0.2 J$), (b) first Brillouin zone. (c) Complete phase diagram in function of the SOC and the external Zeeman field parameters. ($J=1$) }
  \label{phases}
  \end{center}
 \end{figure}

In the two domains in the middle, there is only one topological bandgap characterized by the Chern number of the lower band. At the first transition, the number of edge states passes from 2 to 1, and their group velocities are inverted (due to the sign inversion of the Chern numbers). After the second transition, in the yellow and green domains, the two additional gaps are topological too. Indeed, the presence of topologically protected edge modes is defined by the sum of the Chern numbers of all the bands below the bandgap in consideration ($C=\sum_{i=1} C_n$) \cite{PhysRevLett.71.3697}. By using this argument and looking on the Chern numbers, we can understand that the middle gap becomes trivial in the the fourth phase domain after the band degeneracy occurs and then is lifted at $\Gamma$ point. In the same way, we can see that the two other gaps are still topological. \\

These results show that if $\delta J \neq 0$, there always exists at least one topological gap in the band structure when an external magnetic field is applied. The typical SOC term for exciton-polaritons system is relatively weak and is below $\delta J=0.5$. However, this term cannot vanish since it comes from the TE-TM splitting, a physical phenomenon present in every photonic structure.

\subsection{Rashba spin-orbit coupling}

In the main text, we compare the phase diagrams with TE-TM and Rashba SOCs (Fig. 3(a,b)), with the latter calculated in the linear regime. Here, we present the details of this calculation. Replacing the TE-TM term by the Rashba coupling, we obtain a tight-binding Hamiltonian for electrons in graphene \cite{PhysRevB.79.165442}: 
\begin{eqnarray}
H_k^R=\begin{pmatrix}
\Delta \sigma_Z&G_k \\
G_k^+& \Delta \sigma_Z
\end{pmatrix} 
,\quad G_k=-\begin{pmatrix}
g_kJ&i g_k^+\lambda_R  \\
i g_k^- \lambda_R &g_kJ 
\end{pmatrix}
\end{eqnarray}
The difference with the TE-TM SOC is in the definition of the $g_k^{\pm}$ coefficients, which exhibit a single winding for the Rashba SOC, contrary to the double winding of the TE-TM SOC:
\begin{eqnarray}
  g_k^{\pm}&=& \sum_{j=1}^3 \exp{(-i[\textbf{kd}_{\phi_j}\mp \phi_j])}\\
g_k &=&\sum_{j=1}^3\exp{(-i\textbf{kd}_{\phi_j})} \nonumber
\end{eqnarray}
The total  Hamiltonian is hermitian but we can notice that a $i$ appears in the off-diagonal blocks $G_k$. In electronic systems, the Zeeman term appears generally due to exchange-type interactions \cite{PhysRevB.82.161414} and doesn't need an external magnetic field. In this configuration, the main experimentally tunable parameter is the strength of the Rashba SOC $\lambda_R$, which can be changed by depositing graphene on a substrate or by applying an external electric field.

As an illustration, we plot the spin textures of the first branch of the dispersion in Fig. 6, where panel (a) corresponds to the Rashba SOC, while panel (b) - to the TE-TM SOC. We can see the difference of the in-plane spin winding number around $\Gamma$ ($w_R=1$ for Rashba and $w_{TE-TM}=2$ for TE-TM SOC). However, in both cases the winding numbers around the $K$ and $K'$ points are equal to $1$. 
The topological character of this Hamiltonian with the appearance of the Quantum Anomalous Hall effect, associated with Chern numbers $\pm 2$ due to the additivity of the $K$ and $K'$ contributions to the Berry curvature, have already been suggested  in Ref. \cite{PhysRevB.82.161414}. Moreover, as shown in the main part, a trigonal warping occurs near the $K$ points of the dispersion for Rashba SOC, as well as for the TE-TM SOC. However, we point out that this trigonal warping appears in the $K-\Gamma$ direction for the Rashba SOC, whereas it occurs in the $K-M$ direction for the TE-TM SOC. This difference has important consequences on the topological phase diagram, suppressing one of the phase transitions for the Rashba case: the transition associated with the merging of the additional $K$ points at the $M$ point doesn't exist in the electronic configuration with Rashba SOC.

   \begin{figure}[H]
 \begin{center}
 \includegraphics[scale=0.37]{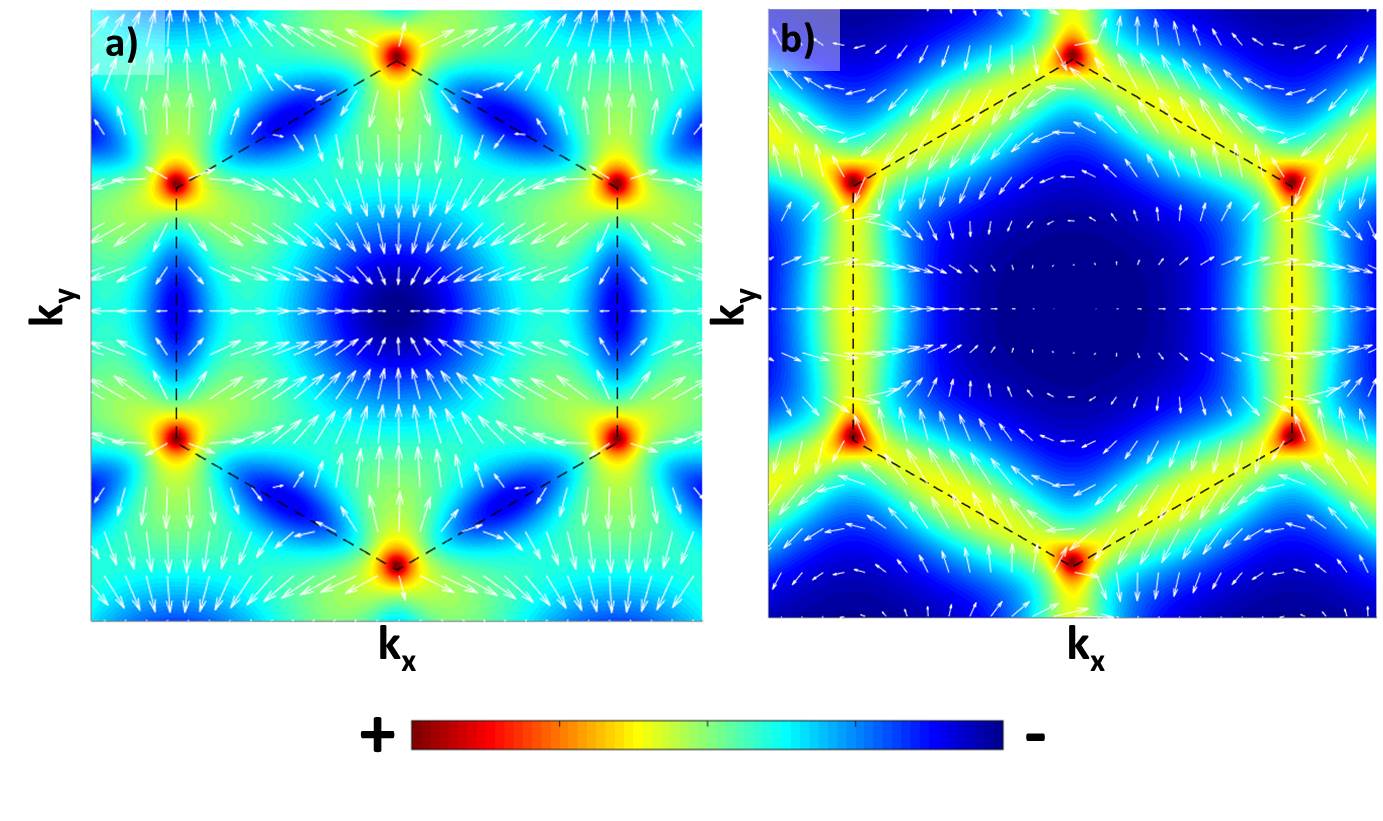}
 \caption{ (Color online) (a)-(b) Polarization textures in presence of Rashba and  TE-TM SOC respectively with $\Delta=0.5$. ($\delta J=\lambda_R=0.2J$). The colors represent the circular components of the spin and the white arrows - the in-plane spin projections.}
  \label{pola}
  \end{center}
 \end{figure}

Hence, by looking on the phase diagram, we can see that the main gap (medium gap) closes only at $2\Delta/J=6$. Below this transition, the main gap is characterized by the presence of 2 edges modes whereas above the transition, the gap becomes trivial.
 
 \section{Bogoliubov matrix}
In this section we present the derivation of the Bogoliubov matrix using the small perturbations approach. This approach consist to write the perturbed macro-occupied state wave function in the following form:
 \begin{equation}
\vec{\Phi}=e^{i(\textbf{k}_p.r-\omega_p t)}(\vec{\Phi}_p+\textbf{u}e^{i(\textbf{k}.r-\omega t)}+\textbf{v}^*e^{-i(\textbf{k}.r-\omega^* t)})
\end{equation}
where $\vec{\Phi}_p=(\Psi_{p,A}^+,\Psi_{p,A}^-,\Psi_{p,B}^+,\Psi_{p,B}^-)^T$ , $\textbf{u}$ and $\textbf{v}$ are vectors of the form $(u_A^+,u_A^-,u_B^+,u_B^-)^T$ too.  
Then, inserting this wave function in the driven dissipative Gross-Pitaevskii equation (main text) and linearizing for  $\textbf{u}$ and $\textbf{v}$, we obtain the following matrix: 
\begin{widetext}
\begin{equation}
\setlength\arraycolsep{0pt}
M=\begin{pmatrix}
(d_A^+-\omega_p) &\alpha_2\Psi_{p,A}^{-*}\Psi_{p,A}^{+}&-f_{k_p+k}J & -f_{k_p+k}^+\delta J & \alpha_1 \Psi_{p,A}^{+2}& \alpha_2 \Psi_{p,A}^{-}\Psi_{p,A}^+&0&0 \\

\alpha_2\Psi_{p,A}^{+*}\Psi_{p,A}^{-} &(d_A^--\omega_p)&-f_{k_p+k}^-\delta J & -f_{k_p+k}J &\alpha_2 \Psi_{p,A}^{-}\Psi_{p,A}^+ & \alpha_1 \Psi_{p,A}^{-2} &0&0\\

 -f_{k_p+k}J &-f_{k_p+k}^{-*}\delta J&(d_B^+-\omega_p)&\alpha_2\Psi_{p,B}^{-*}\Psi_{p,B}^{+}&0&0& \alpha_1 \Psi_{p,B}^{+2}& \alpha_2 \Psi_{p,B}^{-}\Psi_{p,B}^+\\
 
  -f_{k_p+k}^{+*}J &-f_{k_p+k}\delta J& \alpha_2\Psi_{p,B}^{-}\Psi_{p,B}^{+*} &(d_B^--\omega_p)&0&0& \alpha_2 \Psi_{p,B}^{-}\Psi_{p,B}^+&\alpha_1 \Psi_{p,B}^{-2} \\

-\alpha_1 \Psi_{p,A}^{+2*} & -\alpha_2 \Psi_{p,A}^{-*}\Psi_{p,A}^{+*}&0 &0 & (\omega_p-d_A^+)& -\alpha_2\Psi_{p,A}^{-}\Psi_{p,A}^{+*}&f_{k_p-k}^*J&f_{k_p-k}^{+*}\delta J\\

-\alpha_2 \Psi_{p,A}^{-*}\Psi_{p,A}^{+*}&-\alpha_1 \Psi_{p,A}^{-2*} &0 &0& -\alpha_2\Psi_{p,A}^{-*}\Psi_{p,A}^{+} & (\omega_p-d_A^-)&f_{k_p-k}^{-*}\delta J&f_{k_p-k}^{*} J\\

0& 0&-\alpha_1 \Psi_{p,B}^{+2*} &-\alpha_2 \Psi_{p,B}^{-*}\Psi_{p,B}^{+*}&f_{k_p-k}^{*} J & f_{k_p-k}^{-*} \delta J &(\omega_p-d_B^+)&-\alpha_2\Psi_{p,A}^{-}\Psi_{p,A}^{+*} \\

0& 0&-\alpha_2 \Psi_{p,B}^{-*}\Psi_{p,B}^{+*}&-\alpha_1 \Psi_{p,B}^{-2*} &f_{k_p-k}^{+*}\delta J & f_{k_p-k}^{*}  J &-\alpha_2\Psi_{p,B}^{-*}\Psi_{p,B}^{+} & (\omega_p-d_B^-)\\
\end{pmatrix}
\end{equation}
\end{widetext}
The diagonal elements are defined by:
\begin{equation}
d_s^\sigma=2\alpha_1|\Psi_{p,s}^\sigma|^2+\alpha_2|\Psi_{p,s}^{-\sigma}|^2
\end{equation}
The Bogoliubov eigenenergies are finally obtain by diagonalizing this 8 by 8 matrix. 

In the expression for the self-induced field $\Omega_{SI}(\Gamma)=\sqrt{3}\alpha_1n/2$  the factor $1/2$ comes from the presence of two sublattices and the $\sqrt{3}$ appears from resonant pumping, as compared with a blue shift of an equilibrium condensate $\mu=\alpha n$.

\end{document}